\documentclass[a4paper,fleqn,usenatbib]{mnras}

\usepackage[T1]{fontenc}
\usepackage{ae,aecompl}

\usepackage{graphicx}	
\usepackage{amsmath}	
\usepackage{amssymb}	

\title[Transmission spectroscopy of WASP-104b]{An enhanced slope in the transmission spectrum of the hot Jupiter WASP-104b}

\author[G. Chen et al.]
{G.~Chen$^{1}$\thanks{E-mail: guochen@pmo.ac.cn (GC)},
E.~Pall\'{e}$^{2,3}$,
H.~Parviainen$^{2,3}$,
H.~Wang$^{4}$,
R.~van\,Boekel$^{5}$,
\newauthor
F.~Murgas$^{2,3}$,
F.~Yan$^{6}$,
V.\,J.\,S.~B\'ejar$^{2,3}$,
N.~Casasayas-Barris$^{2,3}$,
N.~Crouzet$^{7}$,
\newauthor
E.~Esparza-Borges$^{3}$,
A.~Fukui$^{8,2}$,
Z.~Garai$^{9,10,11}$,
K.~Kawauchi$^{8}$,
S.~Kurita$^{8}$,
\newauthor
N.~Kusakabe$^{13,14}$,
J.\,P.~de\,Leon$^{15}$,
J.~Livingston$^{15}$,
R.~Luque$^{2,3}$,
\newauthor
A.~Madrigal-Aguado$^{2,3}$,
M.~Mori$^{15}$,
N.~Narita$^{16,17,13,2}$,
T.~Nishiumi$^{18,13}$,
\newauthor
M.~Oshagh$^{2,3}$,
M.~S\'{a}nchez-Benavente$^{2,3}$,
M.~Tamura$^{15,13,14}$,
Y.~Terada$^{19,20}$,
\newauthor
N.~Watanabe$^{18,13}$\\
\\
Affiliations are listed at the end of the paper.
}

\date{Accepted 2020 November 11. Received 2020 October 31; in original form 2020 August 20}

\pubyear{2020}

\begin{document}
\label{firstpage}
\pagerange{\pageref{firstpage}--\pageref{lastpage}}
\maketitle

\begin{abstract}
We present the optical transmission spectrum of the hot Jupiter WASP-104b based on one transit observed by the blue and red channels of the DBSP spectrograph at the Palomar 200-inch telescope and 14 transits observed by the MuSCAT2 four-channel imager at the 1.52~m Telescopio Carlos S\'{a}nchez. We also analyse 45 additional {\it K2} transits, after correcting for the flux contamination from a companion star. Together with the transit light curves acquired by DBSP and MuSCAT2, we are able to revise the system parameters and orbital ephemeris, confirming that no transit timing variations exist. Our DBSP and MuSCAT2 combined transmission spectrum reveals an enhanced slope at wavelengths shorter than 630~nm and suggests the presence of a cloud deck at longer wavelengths. While the Bayesian spectral retrieval analyses favour a hazy atmosphere, stellar spot contamination cannot be completely ruled out. Further evidence, from transmission spectroscopy and detailed characterisation of the host star's activity, is required to distinguish the physical origin of the enhanced slope.

\end{abstract}




\section{Introduction}

Transmission spectroscopy \citep{2000ApJ...537..916S} has been working exceptionally well in characterising exoplanet atmospheres since the first atmospheric detection \citep{2002ApJ...568..377C}. Due to the slant viewing geometry, it is a very sensitive technique to detect the opacity sources in the limb of an atmosphere \citep{2005MNRAS.364..649F}, resulting in a variety of atomic, ionic, and molecular detections in exoplanet atmospheres \citep[see the review of][and references therein]{2019ARA&A..57..617M}. 

Recent high-precision observations conducted in both space \citep[e.g.,][]{2016Natur.529...59S,2018AJ....155..156T} and ground \citep[e.g.,][]{2017Natur.549..238S,2018Natur.557..526N,2018A&A...616A.145C,2019AJ....157...21P} start to resolve spectral absorption profiles, enabling us to retrieve chemical abundances and atmospheric metallicity. Empirical population studies are also emerging to disentangle the atmospheric metallicity and elemental ratios imprinted by planet formation histories \citep[e.g.,][]{2014ApJ...793L..27K,2017Sci...356..628W,2019MNRAS.482.1485P,2019ApJ...887L..20W}.

To this end, it becomes necessary to understand the role of clouds/hazes in shaping the spectral signatures of transmission spectrum \citep[e.g.,][]{2016ApJ...817L..16S,2016ApJ...823..109I,2016ApJ...826L..16H,2017AJ....154..261C,2017ApJ...847L..22F,2020NatAs.tmp..114G}. Several hot Jupiters have been observed to exhibit transmission spectra dominated by a slope in the optical wavelengths, for example, HAT-P-12b \citep{2018A&A...620A.142A,2020AJ....159..234W}, HAT-P-18b \citep{2017MNRAS.468.3907K}, HAT-P-32b \citep{2016A&A...590A.100M,2020AJ....160...51A}, HATS-8b \citep{2020AJ....159....7M}, HD 189733b \citep{2011MNRAS.416.1443S,2013MNRAS.432.2917P}, TrES-3b \citep{2016A&A...585A.114P}, WASP-6b \citep{2013ApJ...778..184J,2015MNRAS.447..463N,2020MNRAS.494.5449C}, WASP-12b \citep{2013MNRAS.436.2956S}, WASP-19b \citep{2017Natur.549..238S}, WASP-43b \citep{2020AJ....159...13W}, WASP-69b \citep{2020A&A...641A.158M}. Such a slope could be a potential sign of scattering processes in the planetary atmosphere, which could stem from the H$_2$ molecules, condensates, or photochemical hazes \citep[e.g.,][]{2008A&A...481L..83L,2015A&A...573A.122W,2017MNRAS.464.4247W,2018MNRAS.475...94K,2017MNRAS.471.4355P,2020ApJ...895L..47O}. However, the absorption of metal lines \citep{2020ApJ...898L..14L} and sulfur compounds \citep{2009ApJ...701L..20Z,2018AJ....156..283E}, or the contribution of stellar heterogeneity \citep{2014ApJ...791...55M,2014A&A...568A..99O,2018ApJ...853..122R,2019AJ....157...96R} could also induce spectral signatures mimicking a slope at low spectral resolutions.

Transmission spectroscopy has been conducted mostly in planets of relatively low gravity ($\lesssim$15~m\,s$^{-2}$), because their atmospheres potentially have larger scale heights to produce more signals. Extending to the regime of intermediate gravity (roughly 20--50~m\,s$^{-2}$) can help understand how gravity influences vertical mixing rates, atmospheric composition, and clouds/hazes \citep[e.g.,][]{2019ApJ...881..152K,2020RAA....20..099Z}. Here we report the transmission spectrum for the hot Jupiter WASP-104b ($T_\mathrm{eq}\sim1500$~K), which orbits a G8 star every 1.75 days \citep{2014A&A...570A..64S}. With a mass of $\sim$1.3~M$_\mathrm{Jup}$ and radius of $\sim$1.1~R$_\mathrm{Jup}$, it has a gravity higher than 20~m\,s$^{-2}$. Based on {\it K2} observations, \citet{2018AJ....156...44M} revised the system parameters and found neither signs of transit timing variations (TTVs) nor transit duration variations (TDVs). The {\it K2} phase curve further revealed that WASP-104b is very dark, with a geometric albedo lower than 0.03 at 95\% confidence \citep{2018AJ....156...44M}. \citet{2020AJ....159..137G} presented secondary eclipse observations of WASP-104b by {\it Spitzer}, and reported a dayside brightness temperature of $1716\pm 197$~K in the 3.6~$\mu$m band and $1783\pm 205$~K in the 4.5~$\mu$m band, respectively.  

This paper is organised as follows. In Section \ref{sec:obs}, we present the summary of spectroscopic and photometric observations, and the corresponding data reduction steps. In Section \ref{sec:analysis}, we describe the dilution correction for the flux contamination by a companion star and light-curve modelling processes. In Section \ref{sec:results}, we present the revised system parameters and transmission spectrum, and discuss the observed spectral signatures. Finally, we give our conclusions in Section \ref{sec:conclusion}.

\section{Observations and data reduction}
\label{sec:obs}

\begin{table*}
\centering
\caption{Observation summary.}
\label{tab:obslog}
\begin{tabular}{ccccccccccc}
\hline\hline
\# & Tele. & Instru. & Start night & Start & End & Filter & $t_\mathrm{exp}$ & Airmass$^a$ & Aperture [pix] & Weather$^b$\\
   &       &         &             & UT    & UT  &        & [sec]            &             &\\
\hline
1	& P200	& DBSP	    &  2015-02-13	& 07:24	 & 11:39	& blue, red	& 60, 150	        & 1.20-1.11-1.44	& 20.5, 14 & A\\
\hline
2	& TCS	& MuSCAT2	&  2018-01-12	& 02:24	 & 06:38	& $g, r, i, z_s$	& 3, 2, 3, 4	        & 1.21-1.07-1.29	& 11, 12, 10, 9 & A\\
3	& TCS	& MuSCAT2	&  2018-01-26	& 02:46	 & 06:53	& $g, r, i, z_s$	& 15, 15, 15, 15	& 1.09-1.07-1.68	& 13, 13, 13, 13 & C\\
4	& TCS	& MuSCAT2	&  2018-04-10	& 21:36	 & 01:38	& $g, r, i, z_s$	& 5, 12, 12, 12	        & 1.10-1.07-1.51	& 13, 12, 12, 11 & A\\
5	& TCS	& MuSCAT2	&  2018-04-17	& 21:33	 & 01:18	& $g, r, i, z_s$	& 15, 15, 15, 15	& 1.08-1.07-1.55	& 13, 13, 12, 13 & A\\
6	& TCS	& MuSCAT2	&  2018-12-01	& 03:14	 & 06:01	& $g, i$	        & 10, 10	        & 1.90-1.11-1.11	& 9, 9 & A\\
7	& TCS	& MuSCAT2	&  2018-12-08	& 02:49	 & 06:20	& $r, i, z_s$	& 10, 15, 15	        & 1.87-1.07-1.07	& 12, 11, 9 & B\\
8	& TCS	& MuSCAT2	&  2018-12-15	& 03:25	 & 06:47	& $r, i, z_s$	& 3, 6, 10	        & 1.40-1.07-1.08	& 9, 11, 8 & A\\
9	& TCS	& MuSCAT2	&  2018-12-22	& 03:34	 & 06:46	& $r, i, z_s$	& 15, 30, 40	        & 1.25-1.07-1.11	& 13, 13, 13 & B\\
10	& TCS	& MuSCAT2	&  2019-01-28	& 00:19	 & 04:30	& $r, i, z_s$	& 4, 7, 14	        & 1.46-1.07-1.12	& 12, 12, 11 & B\\
11	& TCS	& MuSCAT2	&  2019-05-17	& 20:44	 & 23:43	& $r, i, z_s$	& 12, 25, 25	        & 1.08-1.08-1.73	& 9, 9, 9 & B\\
12	& TCS	& MuSCAT2	&  2020-02-13	& 22:46	 & 01:46	& $g, r, i, z_s$	& 35, 35, 35, 35	& 1.55-1.09-1.09	& 10, 13, 13, 13 & C\\
13	& TCS	& MuSCAT2	&  2020-03-05	& 23:42	 & 03:43	& $g, r, i, z_s$	& 15, 15, 20, 20	& 1.12-1.07-1.43	& 12, 8, 12, 13 & A\\
14	& TCS	& MuSCAT2	&  2020-03-12	& 00:52	 & 03:51	& $g, r, i, z_s$	& 12, 8, 15, 15	        & 1.08-1.08-1.66	& 10, 11, 9, 9 & B\\
15	& TCS	& MuSCAT2	&  2020-03-21	& 20:10	 & 23:32	& $g, r, i, z_s$	& 12, 8, 15, 15	        & 1.78-1.07-1.07	& 12, 12, 11, 10 & A\\
\hline
\end{tabular}
\begin{flushleft}
{\small \textit{Notes.} 
$^{a}$The first and third values refer to the airmass at the beginning and at the end of the observation. The second value gives the minimum airmass.
$^{b}$Weather conditions: A -- Clear; B --Thin cirrus; C -- Thick cirrus.
}
\end{flushleft}
\end{table*}

\subsection{P200/DBSP spectroscopy}

We observed one transit of WASP-104b on the night of 2015 February 13, using the Double Spectrograph \citep[DBSP;][]{1982PASP...94..586O} installed at the Palomar 200-inch (P200) Hale telescope in California, USA (Program TAP2012B04, PI: G. Chen). A reference star, 2MASS J10422442+0726349, which was $\sim$29 arcsec away, was simultaneously observed with the target star WASP-104. They were placed on a slit of length 128 arcsec and width 10 arcsec. The dichroic D48 was used to split light into the blue and red channels at the wavelength $\sim$4800~\AA. The blue channel employed the 300 lines\,mm$^{-1}$ grating, covering a wavelength range of 3250--4770~\AA~with a dispersion of 2.1~\AA~per pixel. The red channel employed the 158 lines\,mm$^{-1}$ grating, covering a wavelength range of 4600--9850~\AA~with a dispersion of 3.0~\AA~per pixel. The CCDs for both channels have a format of $4096\times 2048$ unbinned pixels, with a pixel scale of 0.389 arcsec\,pix$^{-1}$ for blue and 0.293 arcsec\,pix$^{-1}$ for red. The night was clear, with a median seeing of 2.1 arcsec, varying between 1.5 and 2.7 arcsec. The observation started $\sim$70~min before the transit ingress and ended $\sim$80~min after the transit egress. The exposure times for the blue and red cameras were 60~sec and 150~sec, respectively. The HeNeAr and FeAr arc lamps were observed for the DBSP blue and red channels, respectively, through a 0.5~arcsec narrow slit before and after the science observation, which will be used in the subsequent wavelength calibrations.

\begin{figure}
\centering
\includegraphics[width=\columnwidth]{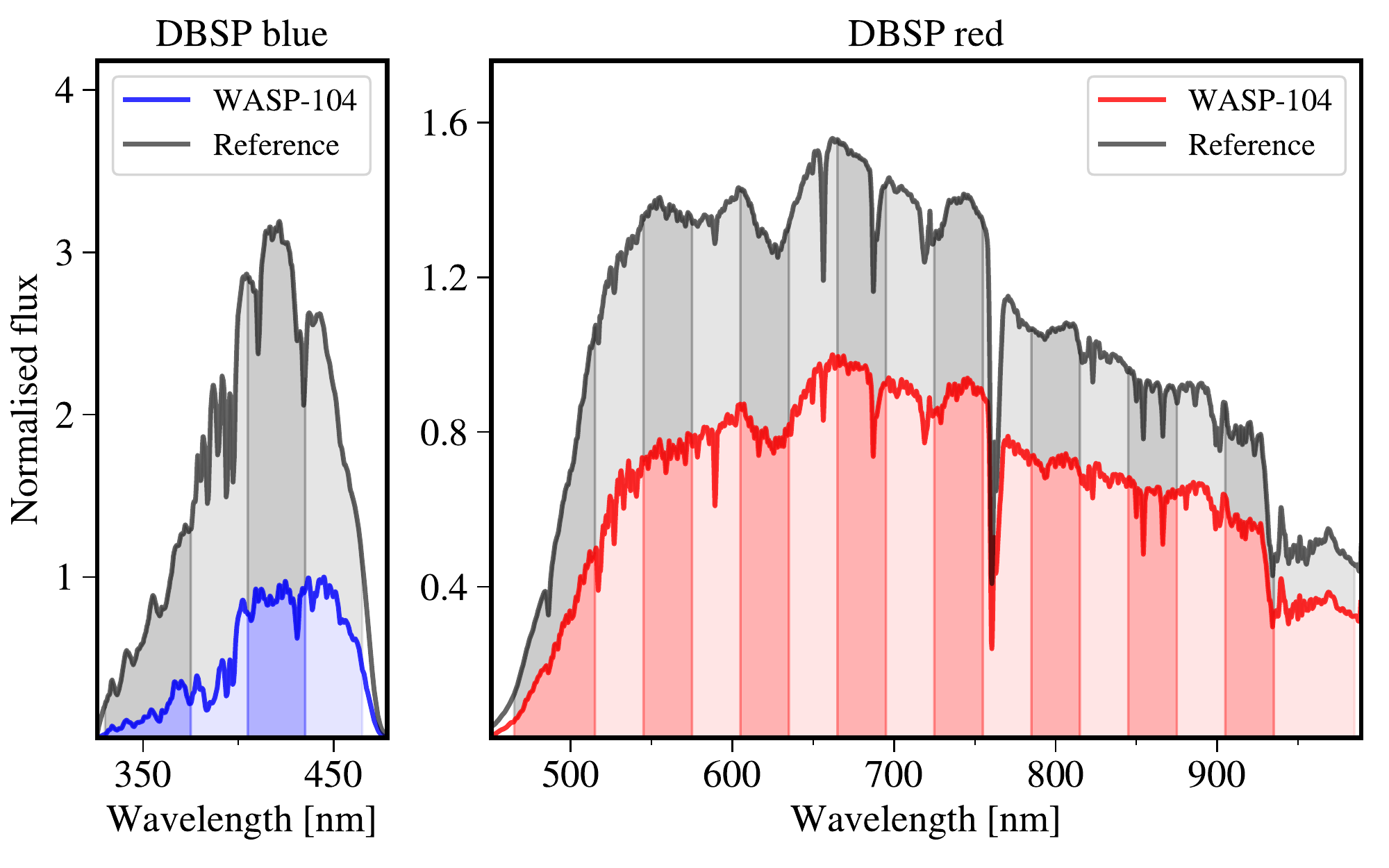}
\caption{Example stellar spectra observed by P200/DBSP. The left and right panels present the DBSP blue and red spectra, respectively. The spectra are normalised to the maximum of WASP-104's spectra. Shaded areas show the spectroscopic passbands. }
\label{fig:dbsp_spec}
\end{figure}

We adapted our GTC data reduction procedures to reduce the acquired P200/DBSP data \citep[e.g.,][]{2017A&A...600L..11C,2017A&A...600A.138C,2018A&A...616A.145C,2020A&A...642A..54C}, which can be generalised to all long-slit data sets. The spectral images were calibrated for overscan, bias, flat, sky background, and cosmic ray hits. One-dimensional spectra were extracted using the optimal extraction algorithm \citep{1986PASP...98..609H}, implemented by the \texttt{APALL} procedure in the \textsc{iraf}\footnote{\url{https://www.iraf.net/}} package \citep{1993ASPC...52..173T}. Different apertures were tested to minimise the scatter of the resulting white light curves. The final spectra were extracted using an aperture radius of 20.5 pixels for the blue channel and 14 pixels for the red channel. The time stamp was constructed in Barycentric Julian Dates in Barycentric Dynamical Time \citep[$\mathrm{BJD}_\mathrm{TDB}$;][]{2010PASP..122..935E}. The spectra's spatial and spectral drifts, and the full widths at half maximum (FWHM) of the point spread function (PSF) were recorded for subsequent light-curve analysis. We created the white light curves within the passband of 330--465~nm for the blue channel, and within 465--985~nm for the red channel. The two channels were further divided into groups of spectroscopic passbands, where it contained four passbands in the blue and 16 in the red, respectively. Most of the spectroscopic passbands have a bin width of 30~nm, except for the passbands on the channel edges. An example of the stellar spectra and spectroscopic passbands are shown in Fig.~\ref{fig:dbsp_spec}.

\subsection{TCS/MuSCAT2 photometry}

We observed 14 transits of WASP-104b using the four-channel imager MuSCAT2 \citep{2019JATIS...5a5001N} installed at the 1.52~m Telescopio Carlos S\'{a}nchez (TCS) in Tenerife, Spain. MuSCAT2 uses three dichroics to split light into four channels, enabling the capability of simultaneous observations in the $g$, $r$, $i$, and $z_s$ bands. It is equipped with four $1024\times 1024$ CCDs to record the images, with a pixel scale of 0.44 arcsec\,pix$^{-1}$, resulting in a field of view of $7.4\times 7.4$~arcmin$^2$. Of the 14 transits, eight were observed in four bands, five were observed in the $r$, $i$, and $z_s$ bands, while one was observed in the $g$ and $i$ bands. This results in a total of 49 light curves. The missing channel in some of the transit observations was due to the replacement of a broken CCD. The $r$ band acquired on the night of 2018 April 10 suffered from saturation issues, and related exposures were discarded. Telluric variations can be well corrected by the reference stars in most nights, except for the night of 2020 February 13, where some of the exposures acquired in the cloud-crossing events had to be discarded. A summary of the MuSCAT2 transit observations is given in Table~\ref{tab:obslog}.

\begin{figure*}
\centering
\includegraphics[width=\textwidth]{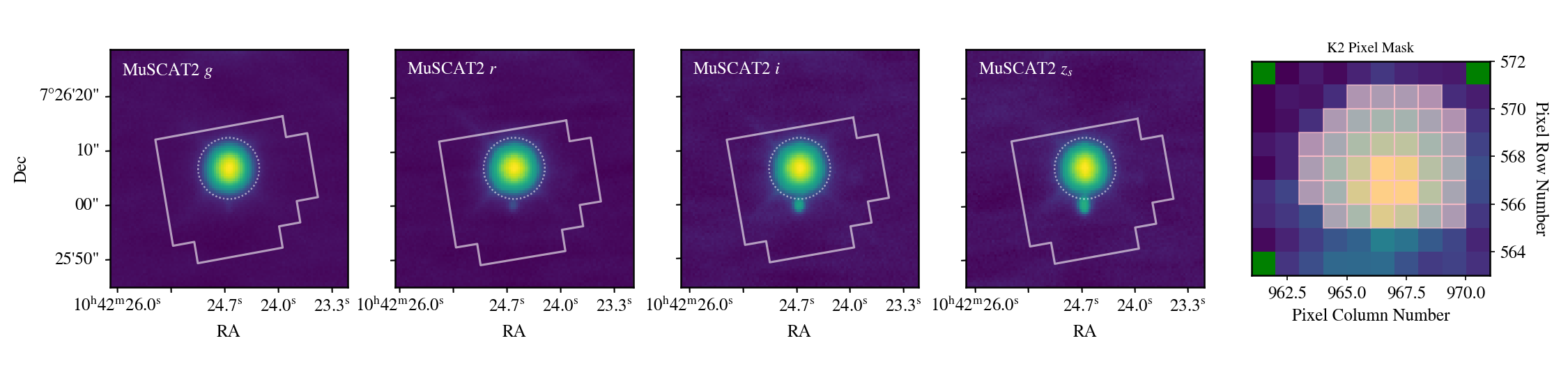}
\caption{{\it K2} custom aperture mask on the four MuSCAT2 bands ({\it left four panels}) and on the {\it K2} target pixel file ({\it right panel}). The dotted circle refers to a radius of 13 MuSCAT2 pixels. The companion star at a separation of 6.84 arcsec can be clearly seen in the $r$, $i$, $z_s$ bands, which is included in the customized {\it K2} aperture mask. We note that the MuSCAT2 and {\it K2} images have different orientations.}
\label{fig:pxmask}
\end{figure*}

We adapted the customized routines outlined in \citet{2014A&A...563A..40C} to reduce the MuSCAT2 data, which was originally designed to reduce the data acquired by another multi-colour imaging instrument GROND \citep{2008PASP..120..405G} that can simultaneously conduct observations in the $g$, $r$, $i$, $z$, $J$, $H$, $K$ bands. The raw images were calibrated for bias and flat. Aperture photometry was performed using the \texttt{APER} routine in the \textsc{idl} \texttt{DAOPHOT}\footnote{\url{https://idlastro.gsfc.nasa.gov/ftp/pro/idlphot/}} package. The centroid was determined by first creating a threshold mask on the stars and then averaging the marginal $x$, $y$ distributions with flux as weights, respectively. Apertures with a grid of increasing radii were tested to minimize the light-curve scatter, with ten different sky annulus sizes being tested for each radius. Meanwhile, different combinations of reference stars were also tested to search for the optimal reference that can minimize the light-curve scatter. The maximum allowed aperture radius was 13 pixels (i.e., 5.72 arcsec), since there is a companion star in the south at a separation of 6.84 arcsec (see Fig.~\ref{fig:pxmask}). The chosen aperture radii for the final photometry are given in Table~\ref{tab:obslog}. The time stamps for the MuSCAT2 observations were also converted to $\mathrm{BJD}_\mathrm{TDB}$, referring to the mid-point of exposures.

\subsection{{\it K2} photometry}

WASP-104 was one of the Campaign 14 targets observed by the {\it K2} mission \citep{2014PASP..126..398H}. The {\it K2} observations spanned 80 days from 2017 May 31 to August 19, covering 45 transits. They have been analysed and published by \citet{2018AJ....156...44M}, however, it is unknown whether the flux contamination by the companion star was corrected or not (see Sect.~\ref{sec:dilution}), and no individual mid-transit times were reported. We therefore downloaded the short cadence target pixel file from Mikulski Archive for Space Telescopes (MAST) using the \textsc{Python} package \texttt{lightkurve}\footnote{\url{https://github.com/KeplerGO/lightkurve}} \citep{2018ascl.soft12013L}. A custom aperture mask (see Fig.~\ref{fig:pxmask}) was created to calculate the Simple Aperture Photometry (SAP) light curve. The systematics were removed using the Pixel Level Decorrelation \citep[PLD; e.g.,][]{2015ApJ...805..132D,2016AJ....152..100L,2018AJ....156...99L} algorithm implemented by \texttt{lightkurve}. Outliers were removed if they were deviating more than 4$\sigma$ from the 100-point Gaussian-convolved light curve. In the end, 115261 data points remained. Since we were only interested in the transit events, we splitted the corrected light curve into 0.15~d pieces centered at the mid-transit time of each individual transit event, which contained a total of 9801 data points.

\section{Analysis}
\label{sec:analysis}

\subsection{Flux contamination from a companion star}
\label{sec:dilution}

As shown in Fig.~\ref{fig:pxmask}, a faint companion star can be noticed toward the south of WASP-104 in the MuSCAT2 images at a distance of 6.84 arcsec, which is 6.65 mag fainter in the {\it Gaia} $G$ band \citep{2018A&A...616A...1G}. The companion star is more prominent in the $z_s$ band, while it is basically invisible in the $g$ band, indicating that it is a late type star. It can be spatially resolved in the ground-based observations, thus its contamination could be eliminated for both DBSP and MuSCAT2 observations. However, it is not possible to separate the companion in the aperture mask of the {\it K2} target pixel file. 

\begin{figure}
\centering
\includegraphics[width=\columnwidth]{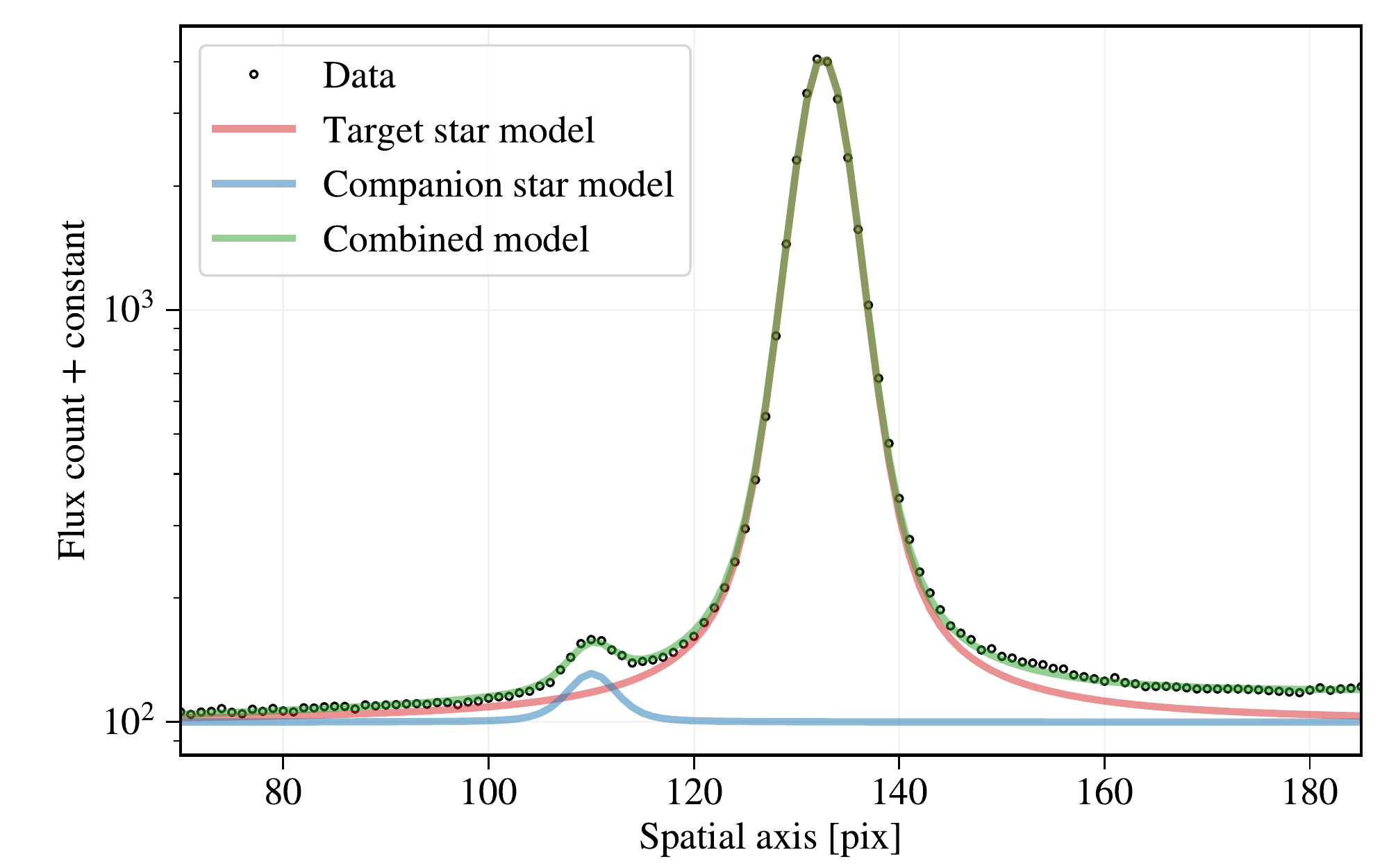}
\includegraphics[width=\columnwidth]{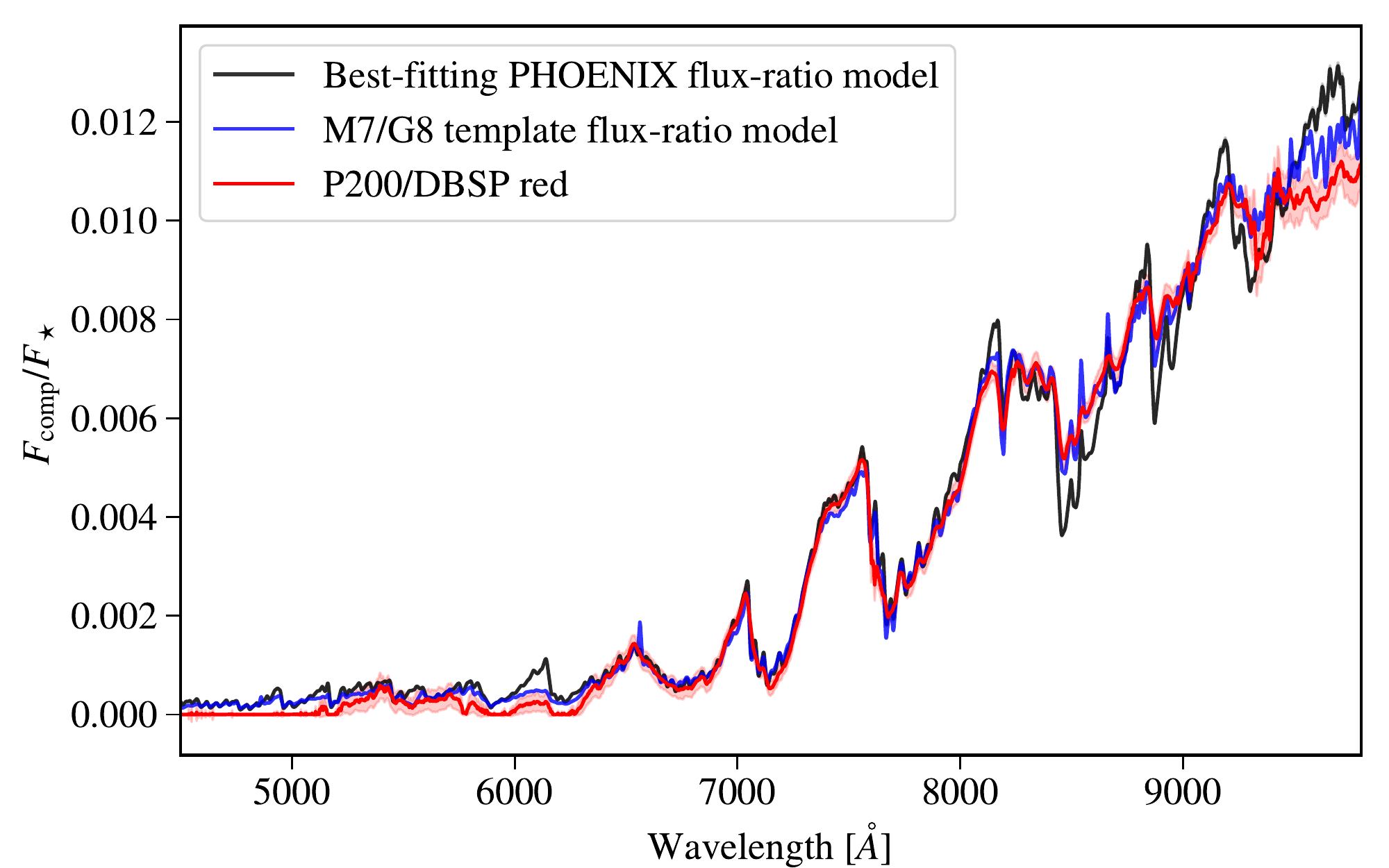}
\caption{{\it Top panel}: an illustration of multi-peak fitting to measure the companion-to-target flux ratio. {Bottom panel}: Flux-ratio spectrum between the companion star and WASP-104. Red line and shaded area show the measured flux ratios derived from P200/DBSP's out-of-transit data. The black line is the best-fitting PHOENIX flux-ratio model. The blue line is the flux-ratio model created from the empirical M7 and G8 spectral templates.}
\label{fig:dilution}
\end{figure}

We attempted to derive the companion-to-target flux ratio ($F_\mathrm{comp}/F_\star$) using the out-of-transit spectra obtained with the DBSP red channel. We used the combination of a Moffat function \citep{1969A&A.....3..455M} and a Lorentz function to fit for the spatial profile. A single peak is described as follows:
\begin{equation}
F=w\times\mathrm{Moffat}+(1-w)\times\mathrm{Lorentz},
\end{equation}
where $F$ is the flux count and $w$ is the weight. The Moffat and Lorentz components share the same centroid and width. A linear trend was fitted together with the target and companion peaks, to correct for the slope introduced by the PSF wing of the reference star 2MASS J10422442+0726349. The fitting was implemented by the \textsc{idl} package \texttt{mpfit}\footnote{\url{https://pages.physics.wisc.edu/~craigm/idl/idl.html}}. An example of the peak fitting is presented in the top panel of Fig.~\ref{fig:dilution}. The derived flux-ratio spectrum is shown in the bottom panel of Fig.~\ref{fig:dilution}. The uncertainties represent the standard deviation of individually measured flux-ratio spectra. We note that the fitting failed at some wavelengths shorter than 6300~\AA, where zero uncertainties were assigned.

We first performed a simple template matching using the empirical library of stellar spectra from the Sloan Digital Sky Survey's Baryon Oscillation Spectroscopic Survey \citep{2017ApJS..230...16K}. An M7V template with a metallicity of $+$0.5 for the companion and a G8V template with a metallicity of $+$0.5 for WASP-104 can best match the measured DBSP flux-ratio spectrum (see the blue line in the bottom panel of Fig.~\ref{fig:dilution}). After integrating in the {\it K2} band using the {\it Kepler} instrument response function\footnote{\url{https://keplergo.arc.nasa.gov/kepler_response_hires1.txt}}, we derived a flux ratio of 0.1737 per~cent.

We then searched for the theoretical spectral templates that can best match the measured flux-ratio spectrum. With certain stellar parameters, e.g., effective temperature $T_\mathrm{eff}$, metallicity $\mathrm{[Fe/H]}$, and surface gravity $\log g$, we interpolated for WASP-104 and the companion star in the grid of spectral templates obtained from the PHOENIX library\footnote{\url{http://phoenix.astro.physik.uni-goettingen.de}} stellar atmospheres \citep{2013A&A...553A...6H}, and convolved the spectra to seeing-limited spectral resolution. The synthetic flux-ratio spectrum was multiplied by a scale factor before it was compared to the data. To fully explore the parameter space, the \textsc{Python} package \texttt{emcee}\footnote{\url{https://github.com/dfm/emcee}} \citep{2013PASP..125..306F} was used to implement the affine-invariant ensemble sampler for Markov chain Monte Carlo (MCMC). During the fitting process, the stellar parameters of WASP-104 were fixed to $T_\mathrm{eff}=5416$~K, $\mathrm{[Fe/H]}=0.40$, and $\log g=4.36$ \citep{2018A&A...620A..58S}, and only the data between 5200 and 9100~\AA~that have successful flux-ratio measurements were used. As a result, we obtained $T_\mathrm{eff}=2827 \pm 6$~K, $\mathrm{[Fe/H]}=0.18\pm 0.03$, and  $\log g=4.46\pm 0.07$ for the companion star, and the scale factor is $f=0.128 \pm 0.002$.

Finally, based on the best-fitting synthetic flux-ratio spectrum (see Fig.~\ref{fig:dilution}), we calculated a flux ratio of $0.1794\pm 0.0075$ per\,cent for the {\it K2} band, which is adopted for the dilution correction in the subsequent analysis. The dilution induced by this flux ratio is very small, i.e., $\Delta R_\mathrm{p}/R_\star=0.000091$ according to our modelling of 45 {\it K2} light curves with or without dilution correction. For completeness, we still corrected it in the light-curve modelling. The DBSP and MuSCAT2 light curves were assumed to be free of flux contamination given the carefully limited aperture radius adopted in photometry.

\subsection{Light-curve modelling}
\label{sec:lcfit}

We have created two broadband light curves from one DBSP transit, 49 broadband light curves from 14 MuSCAT2 transits, and 45 broadband light curves from 45 {\it K2} transits. These amount to a total of 96 broadband light curves. We have also created one set of spectroscopic light curves from the DBSP observation. The broadband light curves will be used to derive common transit parameters such as inclination, semi-major axis, and mid-transit time, while the spectroscopic light curves will be used to derive the transmission spectrum.

In general, the transit model is constructed by the \citet{2002ApJ...580L.171M} analytic model, implemented by the \textsc{Python} package \texttt{batman}\footnote{\url{https://github.com/lkreidberg/batman}} \citep{2015PASP..127.1161K}. The transit model is diluted by a flux ratio of $F_\mathrm{comp}/F_\star$ if the flux contamination exists. The systematics are accounted by Gaussian processes \citep[GP;][]{2006gpml.book.....R,2012MNRAS.419.2683G}, implemented by the \textsc{Python} packages \texttt{george}\footnote{\url{https://github.com/dfm/george}} \citep{2015ITPAM..38..252A}. The covariance between the variables $x_\alpha$ is given by the Mat\'{e}rn $\nu=3/2$ kernel\footnote{The Mat\'{e}rn $\nu=3/2$ kernel is a ``rougher'' generalisation of the squared exponential kernel, which results in smoother functions of the inputs than the latter \citep{2012MNRAS.419.2683G}. We tested both kernels and obtained consistent results, confirming that the kernel choice is not critical for our data sets. Following \citet{2013MNRAS.428.3680G}'s tests, we adopted the Mat\'{e}rn $\nu=3/2$ kernel.} 
\begin{equation}
k_{ij}= A^2\left( 1+\sqrt{3\,D_{ij}^2} \right)\,\exp \left (-\sqrt{3\,D_{ij}^2} \right),
\end{equation}
where $D_{ij}^2=\sum_{\alpha=1}^{N}{(\hat{x}_{\alpha,i}-\hat{x}_{\alpha,j})^2/L_\alpha^2}$, and $A$ and $L_\alpha$ are the characteristic correlation amplitude and length scale. The adopted GP input $x_\alpha$ will be described in the subsections. The transit model is adopted as the GP mean function. A white-noise jitter is added to the photon-noise dominated spectro-/photometric uncertainties in quadrature to account for potential uncertainty underestimation.

The transit model was parameterised by orbital period $P$, orbital inclination $i$, scaled semi-major axis $a/R_\star$, radius ratio $R_\mathrm{p}/R_\star$, mid-transit time $T_\mathrm{C}$, limb-darkening coefficients $u_i$, orbital eccentricity $e$, and argument of periastron $\omega$. The orbital period was fixed to a value of $1.75540636$~d from \citet{2018AJ....156...44M}, while the eccentricity was fixed to zero \citep{2014A&A...570A..64S}. The quadratic limb-darkening law was adopted, and the coefficients were fitted with Gaussian priors $\mathcal{N}(u_i,\sigma^2)$, except for {\it K2}, in which they are freely fitted. The prior mean values $u_i$ were calculated from the ATLAS9 stellar atmosphere models\footnote{\url{http://kurucz.harvard.edu/grids.html}} using a public \textsc{Python} package\footnote{\url{https://github.com/nespinoza/limb-darkening}} \citep{2015MNRAS.450.1879E}, with the stellar parameters of $T_\mathrm{eff}=5500$~K, $\log g=4.5$, and $\mathrm{[Fe/H]}=0.3$. The prior standard deviations $\sigma$ were assumed as the average gap between the $T_\mathrm{eff}=5500$~K, $T_\mathrm{eff}=5250$~K, and $T_\mathrm{eff}=5750$~K grids. The other transit parameters always had uniform priors, while the GP hyperparameters had log uniform priors.

The posterior distributions are explored using MCMC implemented by \texttt{emcee} \citep{2013PASP..125..306F}. Each fitting process always consists of two runs of short chains for the ``burn-in'' phase before a long chain is started. The number of walkers and length of chains depend on how many (hyper-)parameters are fitted and whether or not a convergence can be achieved.

\subsubsection{Fitting of broadband light curves}
\label{sec:fit_wlc}

For some nights in which short exposures (e.g., $\sim$3~sec) were employed, the number of data points in one light curve could be as large as 2300. Given the large number of light curves and data points and different sources of systematics, a global fitting of all the 96 broadband light curves using GP would be computationally expensive. Instead, we adopted the following way to fit for the broadband light curves.

\begin{itemize}
\item {\bf \textit{K2}.} The 45 {\it K2} transit light curves were fitted individually. Time sequence was used as the only input in the GP kernel. Consequently, the free parameters were orbital inclination $i$, scaled semi-major axis $a/R_\star$, radius ratio $R_\mathrm{p}/R_\star$, mid-transit time $T_\mathrm{C}$, limb-darkening coefficients ($u_1$, $u_2$), white noise jitter, and GP hyperparameters ($\ln A$, $\ln L_t$).
\item {\bf MuSCAT2.} The 49 MuSCAT2 transit light curves were first binned in 2~min intervals individually, and then jointly fitted on a nightly basis. As such, it can reduce the computation cost and take advantage of multi colours to mitigate degeneracy in the parameter space of common transit parameters. Multi-dimensional inputs were used for the GP kernel, including time sequence, target star's centroid $x$ and $y$, and PSF's FWHM $s$. For each night, the multi-colour light curves were jointly fitted, sharing the same $i$, $a/R_\star$, and $T_\mathrm{C}$. The other transit parameters ($R_\mathrm{p}/R_\star$, $u_1$, $u_2$), white noise jitter, and GP hyperparameters ($\ln A$, $\ln L_t$, $\ln L_x$, $\ln L_y$, $\ln L_s$) were light curve dependent. 
\item {\bf DBSP.} The two broadband light curves of the DBSP blue and red channels were jointly fitted. The same multi-dimensional GP inputs as those of MuSCAT2 were used. The two light curves shared the same $i$, $a/R_\star$, and $T_\mathrm{C}$, but different $R_\mathrm{p}/R_\star$, limb-darkening coefficients ($u_1$, $u_2$), white noise jitter, and GP hyperparaameters ($\ln A$, $\ln L_t$, $\ln L_x$, $\ln L_y$, $\ln L_s$). 
\end{itemize}

\begin{figure}
\centering
\includegraphics[width=\columnwidth]{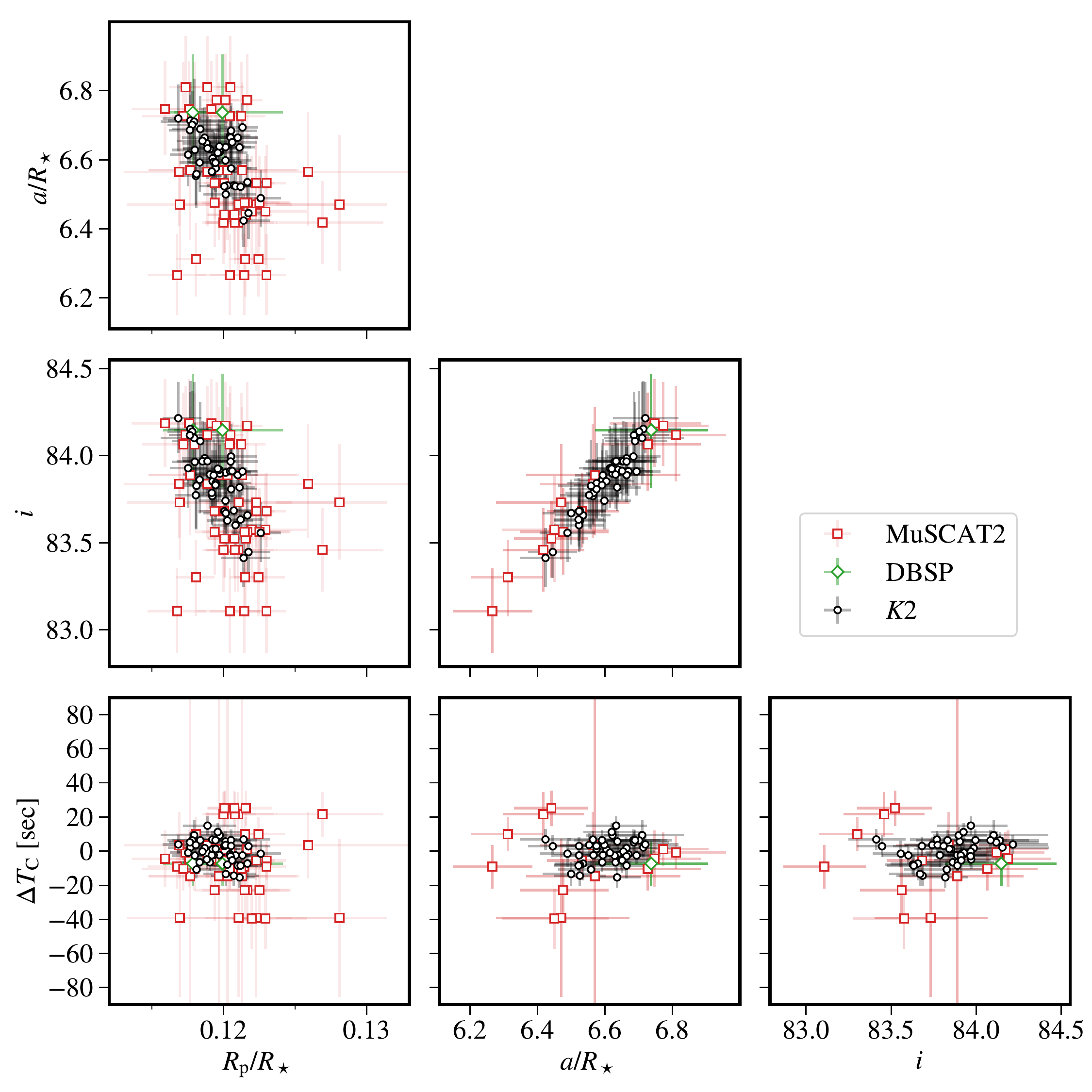}
\includegraphics[width=\columnwidth]{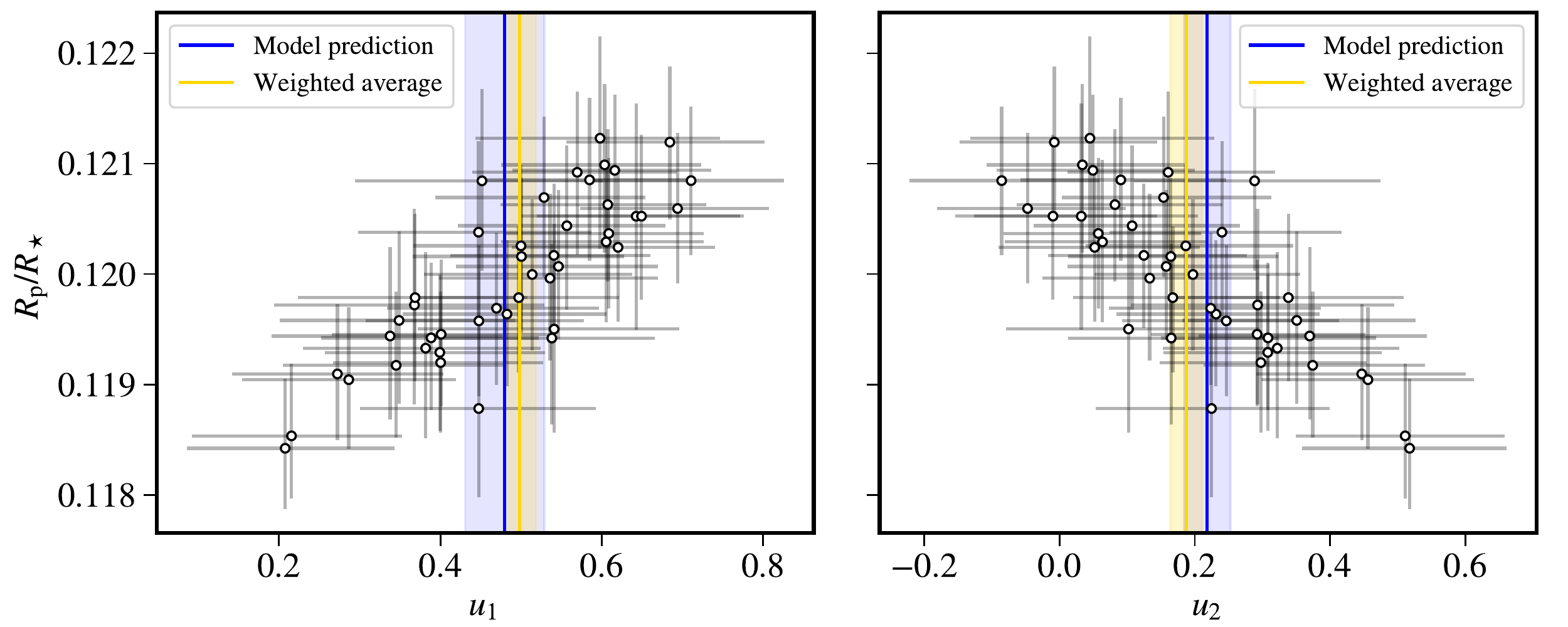}
\caption{Correlation between pairs of transit parameters measured in individual light curves. Red squares, green diamond, and black circles refer to the measurements from MuSCAT2, DBSP, and {\it K2}, respectively.}
\label{fig:par_corr}
\end{figure}

\begin{figure}
\centering
\includegraphics[width=\columnwidth]{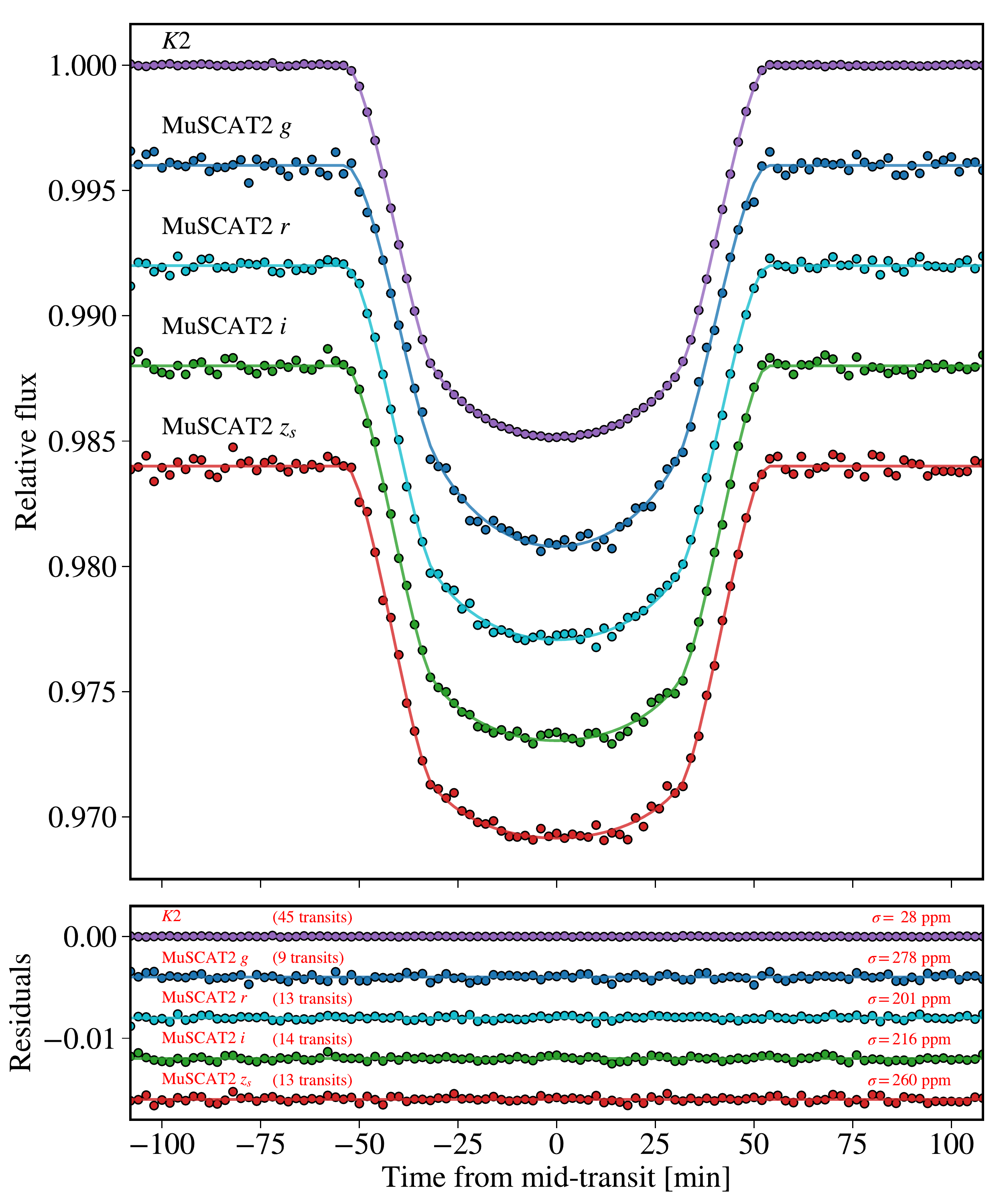}
\caption{Phase-folded {\it K2} and MuSCAT2 light curves, where systematics have been corrected. The light curves have been added constant offsets and binned in 2~min intervals for display clarify.}
\label{fig:k2_lc}
\end{figure}

From the described analyses, we derived a group of $i$, $a/R_\star$, $R_\mathrm{p}/R_\star$, and $\Delta T_\mathrm{C}$, where $\Delta T_\mathrm{C}$ is the deviation from the linear ephemeris based $T_\mathrm{C}$. Unsurprisingly, as shown in Fig.~\ref{fig:par_corr}, a clear correlation exists between $i$ and $a/R_\star$. Correlation can also be observed between $R_\mathrm{p}/R_\star$ and $i$, or between $R_\mathrm{p}/R_\star$ and $a/R_\star$, while they are not evident between $\Delta T_\mathrm{C}$ and other transit parameters. The correlations are more evident for the {\it K2} measurements due to higher photometric precision. The {\it K2} limb-darkening coefficients cannot be constrained in a single transit and exhibit clear correlation with $R_\mathrm{p}/R_\star$. However, the weighted average values of all 45 transits are consistent with the predictions based on stellar atmosphere models. This indicates that the Gaussian priors imposed on the MuSCAT2 and DBSP data are reasonable. We calculated the weighted average values for $i$ and $a/R_\star$, and list them in Table~\ref{tab:avg_par}. The MuSCAT2 and DBSP weighted average values agree with those of {\it K2} within 2$\sigma$. 

To eliminate the correlation's impact on the derived wavelength dependent radius ratios, we performed the light-curve modelling again for the {\it K2}, MuSCAT2, and DBSP light curves. For this time, the values of $i$ and $a/R_\star$ were fixed to those weighted average values listed in Table \ref{tab:avg_par}, and $T_\mathrm{C}$ was fixed to those determined in the previous step (see Table~\ref{tab:tmid}). Since $i$ and $a/R_\star$ were fixed to the same values, the light curves were fitted individually for either {\it K2}, or MuSCAT2, or DBSP blue and red. We note that the original unbinned MuSCAT2 light curves were used for this particular analysis. Finally, for each band, the radius ratio values were weighted averaged, and given in Table \ref{tab:finalpar}.

\begin{figure*}
\centering
\includegraphics[width=\textwidth]{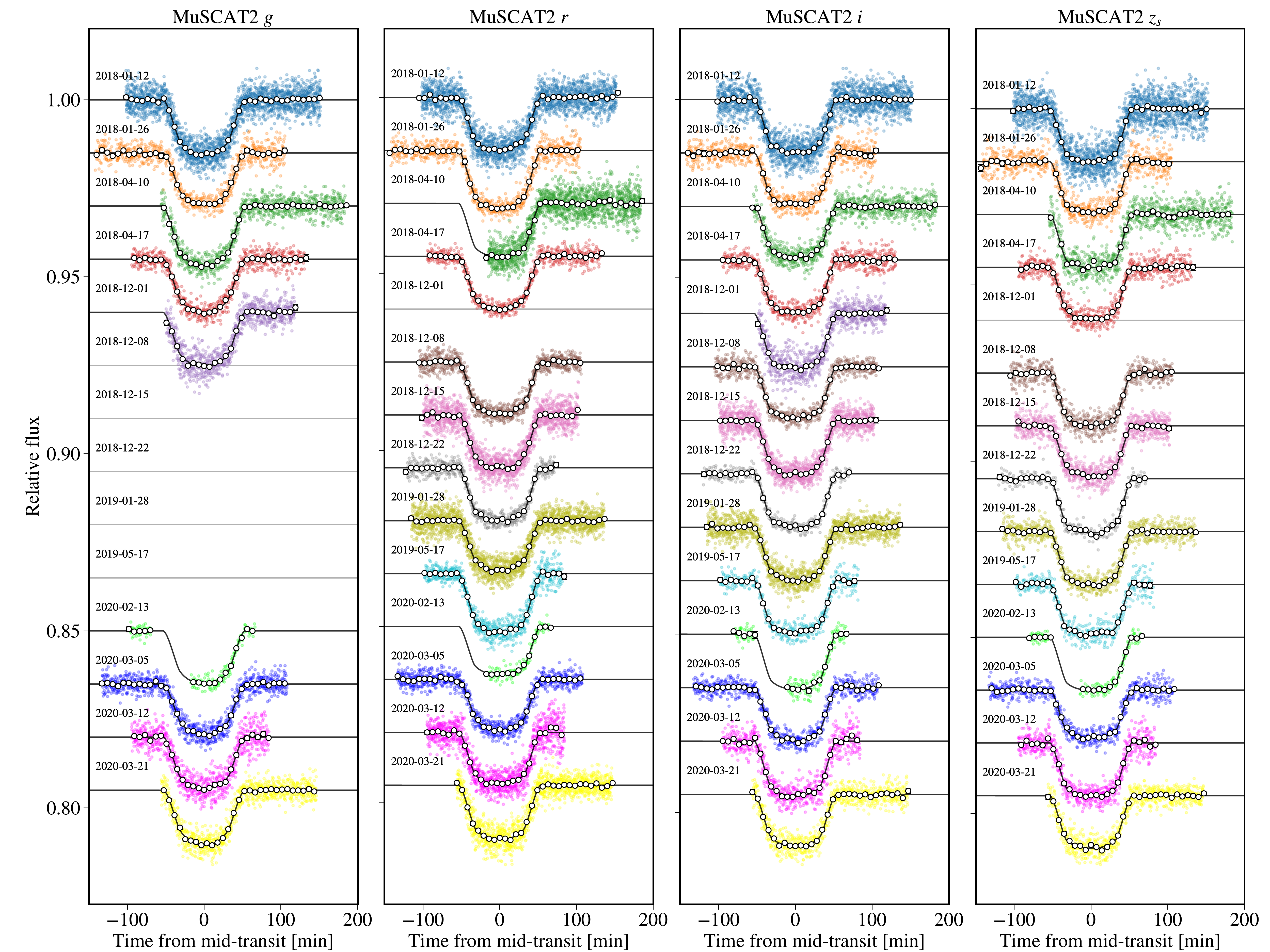}
\caption{Multi-colour light curves obtained with TCS/MuSCAT2 on 14 nights. From {\it left} to {\it right} shows the systematics corrected light curves in the $g$, $r$, $i$, $z_s$ bands, respectively. The colour dots refer to the original unbinned exposures, while the black circles refer to light curves binned in 7 min intervals for display purpose.}
\label{fig:m2_lc}
\end{figure*}

\begin{figure*}
\centering
\includegraphics[width=\textwidth]{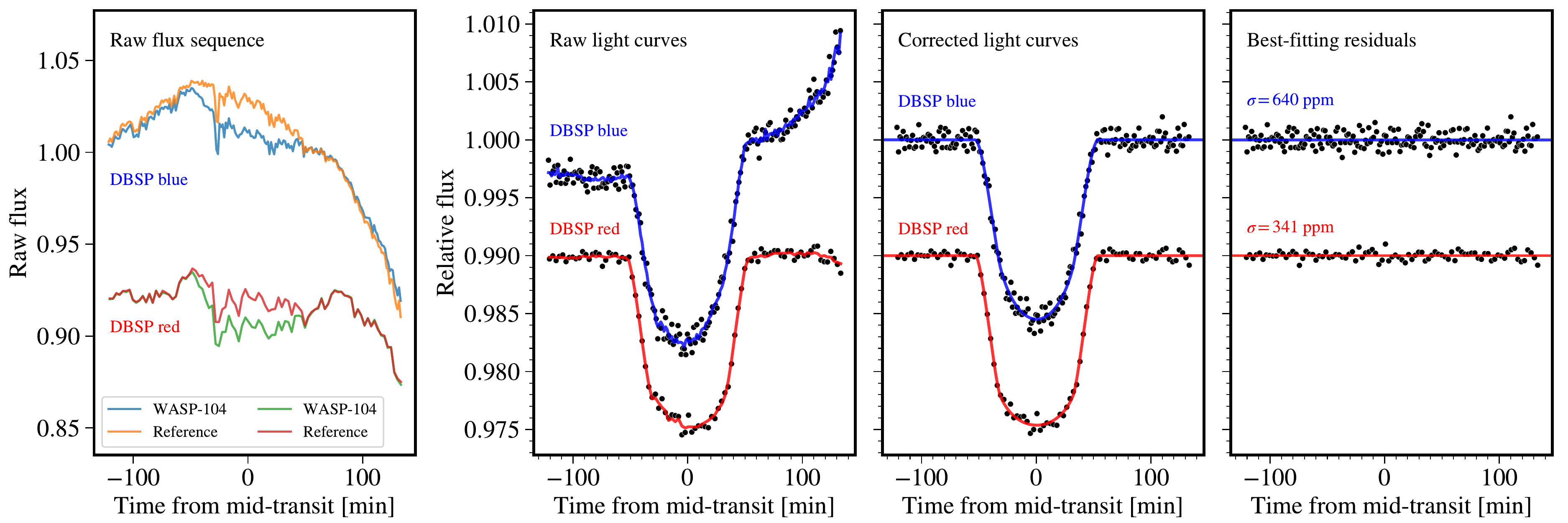}
\caption{Broadband light curves obtained with the P200/DBSP blue and red channels on the night of 2015 February 13. The {\it left panel} presents the raw flux sequences of the target and reference stars. The three {\it right panels} present the reference-calibrated light curves: before systematics correction, after systematics correction, and best-fitting residuals.}
\label{fig:dbsp_wlc}
\end{figure*}

The phase-folded {\it K2} and MuSCAT2 light curves are shown in Fig.~\ref{fig:k2_lc}. The 49 unbinned MuSCAT2 light curves are presented in Fig.~\ref{fig:m2_lc}. The two DBSP broadband light curves are presented in Fig.~\ref{fig:dbsp_wlc}. From the modelling of the broadband light curves, the standard deviation of unbinned light-curve residuals is 1.00--2.43$\times$ photon noise for MuSCAT2, 2.12$\times$ photon noise for DBSP blue, 5.81$\times$ photon noise for DBSP red, and 0.85--1.01$\times$ photon noise for {\it K2}, respectively. When phase-folded into 2 min intervals, the light-curve residuals show an RMS of 28~ppm for {\it K2} (45 transits), 278~ppm for MuSCAT2 $g$ (9 transits), 201~ppm for MuSCAT2 $r$ (13 transits), 216~ppm for MuSCAT2 $i$ (14 transits), and 260~ppm for MuSCAT2 $z_s$ (13 transits).

\begin{table}
\centering
\caption{Common transit parameters derived from this work.}
\label{tab:avg_par}
\begin{tabular}{lcc}
\hline\hline
Case  & $i$ [deg] & $a/R_\star$\\
\hline
{\it K2}+MuSCAT2+DBSP    & $83.811 \pm 0.025$ & $6.591 \pm 0.012$ \\
({\bf adopted}) & & \\
\hline
{\it K2}                 & $83.827 \pm 0.027$ & $6.599 \pm 0.012$ \\
MuSCAT2+DBSP             & $83.713 \pm 0.068$ & $6.527 \pm 0.035$ \\
\hline
\end{tabular}
\end{table}

\begin{table}
\centering
\caption{Derived transit parameters for the WASP-104 system.}
\label{tab:finalpar}
\begin{tabular}{lr}
\hline\hline
Parameter & Value \\
\hline
Radius ratio, $R_\mathrm{p}/R_\star$$^a$      &   \\
\hspace{1em} {\it K2}      & $0.11992 \pm 0.00011$ \\\noalign{\smallskip}
\hspace{1em} MuSCAT2 $g$   & $0.12098 \pm 0.00046$ \\\noalign{\smallskip}
\hspace{1em} MuSCAT2 $r$   & $0.12002 \pm 0.00030$ \\\noalign{\smallskip}
\hspace{1em} MuSCAT2 $i$   & $0.12031 \pm 0.00026$ \\\noalign{\smallskip}
\hspace{1em} MuSCAT2 $z_s$ & $0.11998 \pm 0.00033$ \\\noalign{\smallskip}
\hspace{1em} DBSP blue     & $0.1222  ^{+0.0035}_{-0.0034}$ \\\noalign{\smallskip}
\hspace{1em} DBSP red      & $0.11896 ^{+0.00073}_{-0.00084}$ \\\noalign{\smallskip}
Orbital inclination, $i$ [deg]         & $83.811 \pm 0.025$  \\\noalign{\smallskip}
Scaled semi-major axis, $a/R_\star$    & $6.591 \pm 0.012$  \\\noalign{\smallskip}
Transit duration, $T_{14}$ [d]         & $0.074109 \pm 0.000063$  \\\noalign{\smallskip}
Ingress(egress) duration, $T_{12}$ [d] & $0.014907 \pm 0.000071$  \\\noalign{\smallskip}
Impact parameter, $b$                  & $0.7147 \pm 0.0014$  \\
\hline
Transit epoch, $T_0$ [MJD$^{b}$]       & $7935.070228 \pm 0.000011$ \\\noalign{\smallskip}
Orbital period, $P$ [d]                & $1.75540563 \pm 0.00000011$ \\
\hline
\end{tabular}
\begin{flushleft}
{\color{black}\small \textit{Notes.} 
$^{a}$The presented values of $R_\mathrm{p}/R_\star$ are weighted average of each band, derived from the analyses where $i$, $a/R_\star$, and $T_\mathrm{C}$ were fixed. 
$^{b}$$\mathrm{MJD}=\mathrm{BJD}_\mathrm{TDB}-2450000$. 
}
\end{flushleft}
\end{table}

\begin{figure*}
\centering
\includegraphics[width=\textwidth]{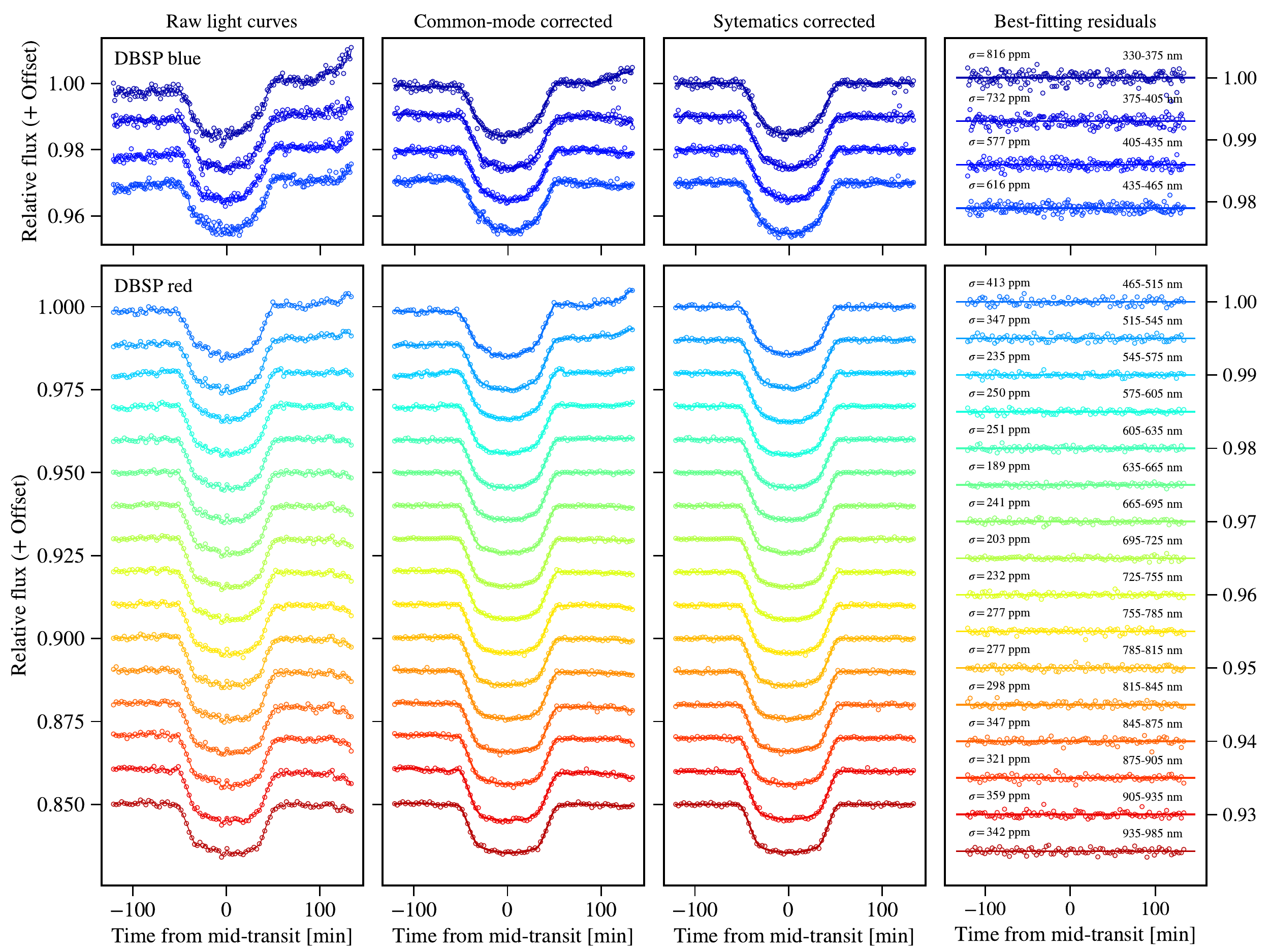}
\caption{Spectroscopic light curves obtained with P200/DBSP on the night of 2015 February 13. The two rows show the light curves from the DBSP blue and red channels, respectively. From {\it left} to {\it right} presents the reference-calibrated raw light curves, common-mode noise corrected light curves, systematics corrected light curves, and best-fitting residuals, respectively.}
\label{fig:dbsp_slc}
\end{figure*}

\begin{figure}
\centering
\includegraphics[width=\columnwidth]{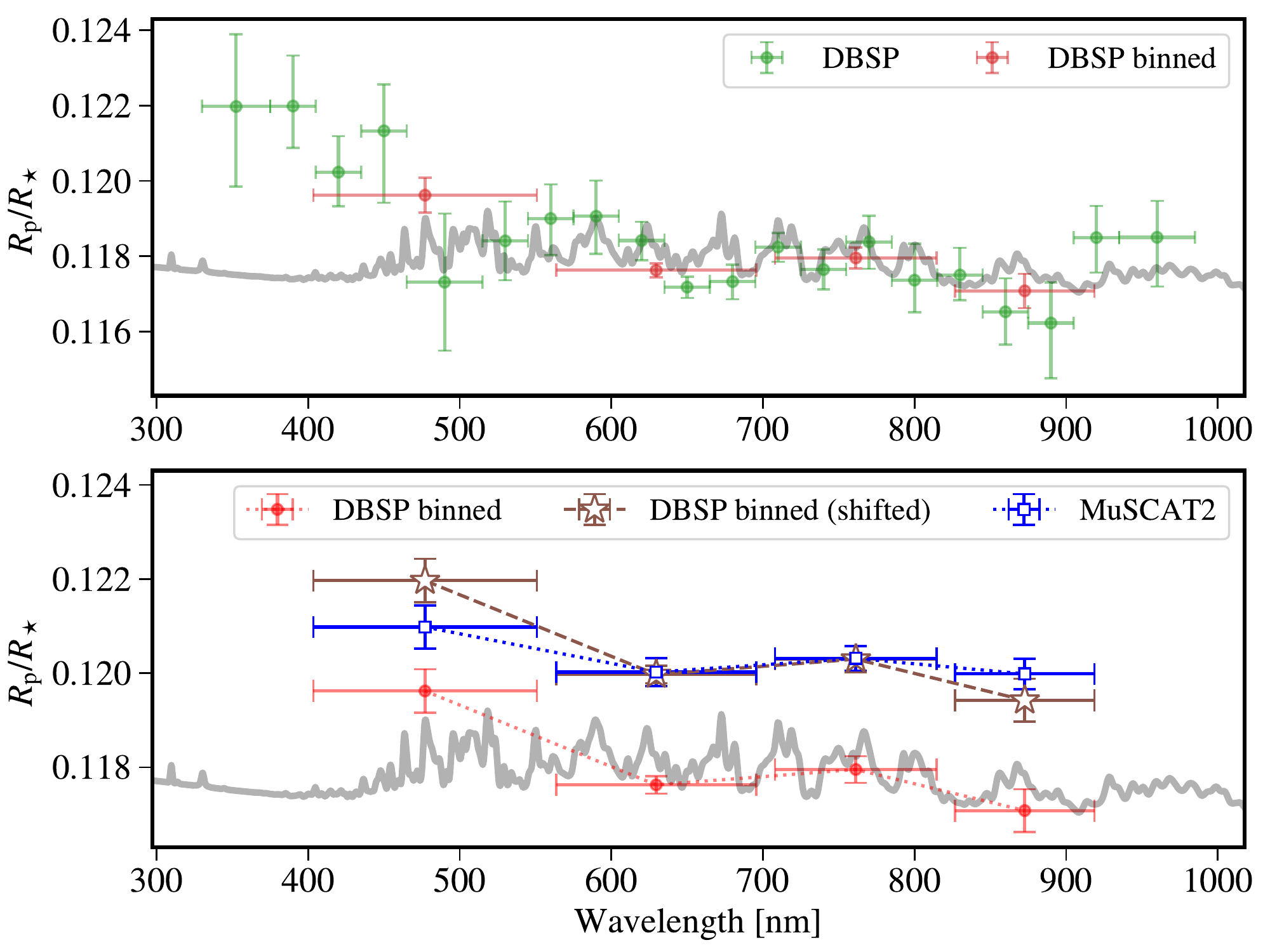}
\caption{{\it Top panel:} Original DBSP transmission spectrum (in green) compared to the one binned into the MuSCAT2 bandpasses (in red). A 1$\times$solar cloud-free model is also shown, calculated by \texttt{PLATON} (see Sect.~\ref{sec:retrieval_atm}) with WASP-104b's physical parameters. {\it Bottom panel:} Comparison between MuSCAT2 and DBSP, where the latter has been shifted upwards by $\Delta R_\mathrm{p}/R_\star=0.00234$ (in brown). The MuSCAT2 data are shown in blue.}
\label{fig:dbsp_cmp}
\end{figure}

\subsubsection{Fitting of spectroscopic light curves}

\begin{table}
\centering
\caption{Transmission spectrum measured by P200/DBSP.}
\label{tab:dbsp_transpec}
\begin{tabular}{ccc}
\hline\hline
 \# & Wavelength (nm) & $R_\mathrm{p}/R_\star$ \\
\hline
  1 & 330 -- 375 & $0.12198 ^{+0.00192}_{-0.00213}$\\\noalign{\smallskip}
  2 & 375 -- 405 & $0.12199 ^{+0.00134}_{-0.00111}$\\\noalign{\smallskip}
  3 & 405 -- 435 & $0.12023 ^{+0.00096}_{-0.00090}$\\\noalign{\smallskip}
  4 & 435 -- 465 & $0.12133 ^{+0.00124}_{-0.00191}$\\\noalign{\smallskip}
  5 & 465 -- 515 & $0.11731 ^{+0.00182}_{-0.00182}$\\\noalign{\smallskip}
  6 & 515 -- 545 & $0.11841 ^{+0.00105}_{-0.00105}$\\\noalign{\smallskip}
  7 & 545 -- 575 & $0.11900 ^{+0.00090}_{-0.00097}$\\\noalign{\smallskip}
  8 & 575 -- 605 & $0.11906 ^{+0.00094}_{-0.00100}$\\\noalign{\smallskip}
  9 & 605 -- 635 & $0.11842 ^{+0.00050}_{-0.00052}$\\\noalign{\smallskip}
 10 & 635 -- 665 & $0.11718 ^{+0.00028}_{-0.00029}$\\\noalign{\smallskip}
 11 & 665 -- 695 & $0.11733 ^{+0.00045}_{-0.00047}$\\\noalign{\smallskip}
 12 & 695 -- 725 & $0.11824 ^{+0.00037}_{-0.00039}$\\\noalign{\smallskip}
 13 & 725 -- 755 & $0.11765 ^{+0.00053}_{-0.00053}$\\\noalign{\smallskip}
 14 & 755 -- 785 & $0.11838 ^{+0.00069}_{-0.00072}$\\\noalign{\smallskip}
 15 & 785 -- 815 & $0.11736 ^{+0.00097}_{-0.00085}$\\\noalign{\smallskip}
 16 & 815 -- 845 & $0.11750 ^{+0.00072}_{-0.00067}$\\\noalign{\smallskip}
 17 & 845 -- 875 & $0.11652 ^{+0.00089}_{-0.00087}$\\\noalign{\smallskip}
 18 & 875 -- 905 & $0.11622 ^{+0.00108}_{-0.00146}$\\\noalign{\smallskip}
 19 & 905 -- 935 & $0.11849 ^{+0.00084}_{-0.00093}$\\\noalign{\smallskip}
 20 & 935 -- 985 & $0.11850 ^{+0.00096}_{-0.00130}$\\\noalign{\smallskip}
\hline
\end{tabular}
\end{table}

To derive the DBSP transmission spectrum for WASP-104b, we performed light-curve modelling on the DBSP spectroscopic light curves. The transit parameters $i$ and $a/R_\star$ were fixed to the weighted average values listed in Table~\ref{tab:avg_par}. The mid-transit time was fixed to the one derived from the DBSP joint fitting as described in Sect.~\ref{sec:fit_wlc} (see Table~\ref{tab:tmid}). 

We tested various state vectors as the GP inputs, individually or in combination, such as time sequence $t$, spectra's drift in the spectral and spatial directions ($x,y$), and PSF's FWHM $s$, and calculated corresponding Bayesian information criterion \citep[BIC;][]{schwarz1978} and Akaike information criterion \citep[AIC;][]{1974ITAC...19..716A}. The combination $(t,x,y,s)$ resulted in the lowest AIC value, while $(t,y,s)$ resulted in the lowest BIC value. The transmission spectra derived from these two GP input combinations exhibit almost no difference ($\chi^2=0.05$ for 20 degrees of freedom, hereafter dof). We also tested using only the transit model as the GP mean function, as well as multiplying the transit model by a baseline trend. The results were consistent, but the latter exhibited higher AIC and BIC values. For our final choice, we adopted the transit model as the GP mean function, and $(t,y,s)$  as the GP inputs. The free parameters were radius ratio $R_\mathrm{p}/R_\star$, limb-darkening coefficients ($u_1$,$u_2$), white noise jitter, and GP hyperparameters ($\ln A$, $\ln L_t$, $\ln L_y$, $\ln L_s$). 

We first performed the light-curve modelling directly on the original DBSP spectroscopic light curves. We then derived the common-mode noise in a way slightly different from previous studies \citep[e.g.,][]{2018A&A...616A.145C,2020A&A...642A..54C,2019A&A...622A.172M,2020A&A...641A.158M}, due to strong flux variation across wavelengths. The spectroscopic light curves were divided by their best-fitting transit model individually. The resulting light-curve systematics residuals were averaged for the DBSP blue and red channels, respectively. The average common-mode systematics empirical model was then corrected in the original spectroscopic light curves. The common-mode corrected spectroscopic light curves were fitted to derive the final transmission spectrum. The transmission spectrum derived from the common-mode corrected data set is consistent with the one derived from the original data set ($\chi^2=8.0$ for 20 dof), but with smaller error bars because the achromatic common-mode systematics trend has been removed.

The DBSP spectroscopic light curves are presented in Fig.~\ref{fig:dbsp_slc}. The derived DBSP transmission spectrum is given in Table~\ref{tab:dbsp_transpec}. For the DBSP spectroscopic light curves, the standard deviation of light-curve residuals is 1.2--2.2$\times$ photon noise, and it decreases to 0.9--1.4$\times$ photon noise after removing the common-mode noise.

In addition to DBSP, we have derived a broadband transmission spectrum from the MuSCAT2 multi-colour light curves in the second step described in Sect.~\ref{sec:fit_wlc}. As can be seen in Fig.~\ref{fig:dbsp_cmp}, the spectral shape of the MuSCAT2 broadband transmission spectrum agrees well with DBSP when the latter is binned into the same bandpasses. However, there is a constant offset of $\Delta R_\mathrm{p}/R_\star=0.00234\pm0.00022$ between these two transmission spectra.

\section{Results and discussion}
\label{sec:results}

\subsection{Revised system parameters}

\subsubsection{Transit parameters and TTVs}

\begin{figure*}
\centering
\includegraphics[width=\textwidth]{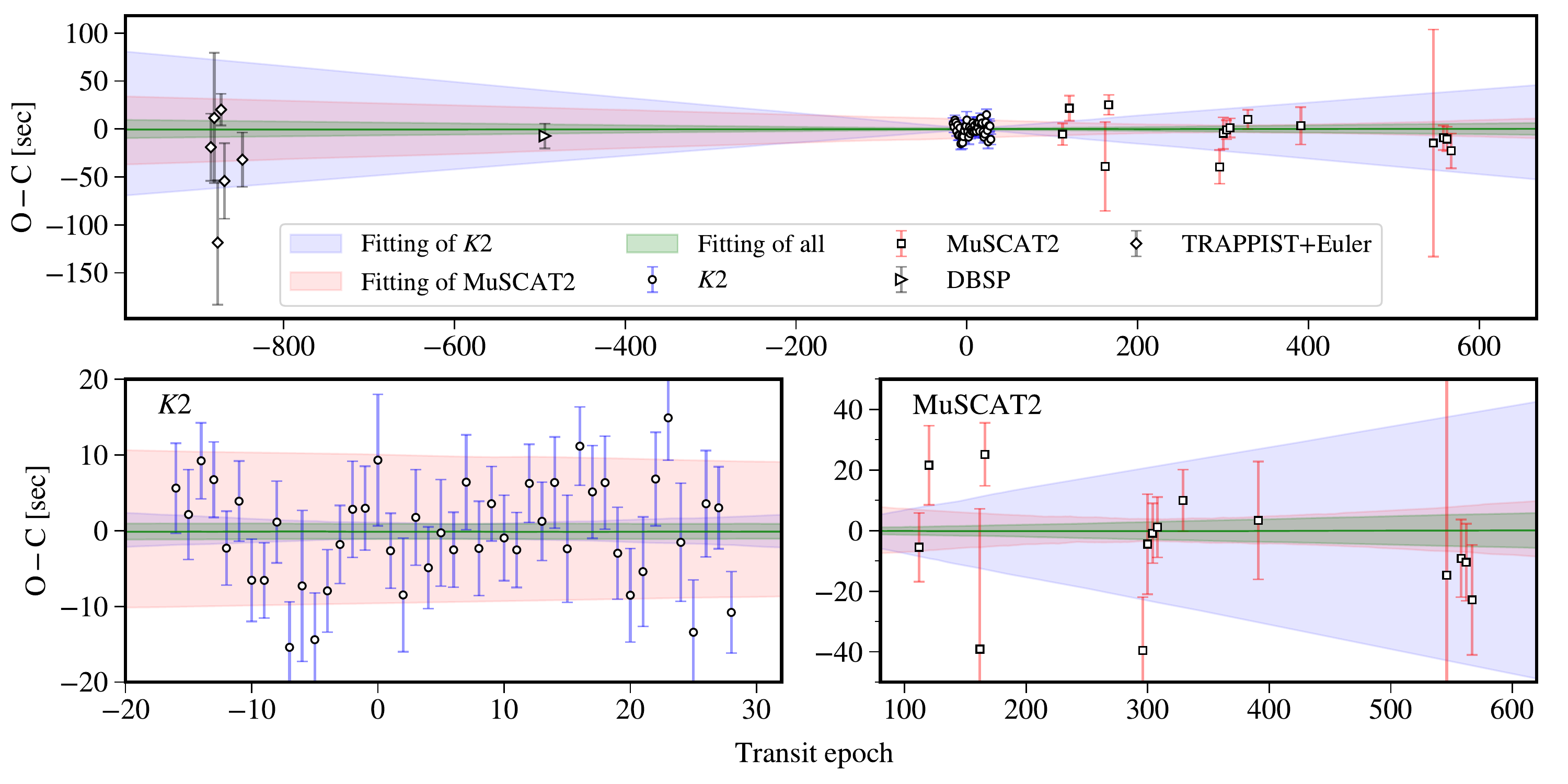}
\caption{Residuals of the mid-transit times after a linear ephemeris regression. {\it Top panel} shows all the mid-transit times derived from this work, which include the newly observed DBSP and MuSCAT2 data, and the re-analysed {\it K2}, TRAPPIST, and Euler data \citep{2014A&A...570A..64S,2018AJ....156...44M}. {\it Bottom panel} shows the zoom-in views on the {\it K2} ({\it bottom left}) and MuSCAT2 ({\it bottom right}) measurements. The green shaded area present the 1$\sigma$ confidence region of the fitting on all the available data. In contrast, the blue and red shaded areas present the 1$\sigma$ confidence regions of the fittings on the {\it K2}-only and MuSCAT2-only data, respectively.}
\label{fig:ephemeris}
\end{figure*}

Based on the newly collected MuSCAT2 and DBSP light curves and the re-analysed {\it K2} light curves, we are able to revise the transit parameters for the WASP-104 planetary system, which are presented in Table~\ref{tab:finalpar}. 

As shown in Table~\ref{tab:avg_par}, we have obtained consistent values for $i$ and $a/R_\star$ between {\it K2} and MuSCAT2+DBSP. However, our inclination $i=83.811^{\circ}\pm 0.025^{\circ}$ and impact parameter $b=0.7147 \pm 0.0014$ are deviating more than 5$\sigma$ from those ($83.612^{\circ}\pm 0.026^{\circ}$, $0.7278 \pm 0.0016$) of \citet{2018AJ....156...44M}. Therefore, we also recalculated the transit duration $T_{14}=0.074109 \pm 0.000063$~d and the ingress (or egress) duration $T_{12}=0.014907 \pm 0.000071$~d using the equations (2) and (3) from \citet{2003ApJ...585.1038S}. Again, the differences are significant ($\Delta T_{14}=1.85\pm 0.11$~min and $\Delta T_{12}=-0.66\pm 0.15$~min) from those of \citet{2018AJ....156...44M}. Given that we have more light curves observed in different colours, we report our newly revised parameters.

To revise the transit ephemeris, we re-measured the mid-transit times for four TRAPPIST and two Euler transit light curves from \citet{2014A&A...570A..64S}. The light curves were fitted individually, and the GP was implemented by \texttt{george} with the Mat\'{e}rn $\nu=3/2$ kernel, where only the time sequence was adopted as the GP input. Consequently, there are a total of 66 unique mid-transit time measurements, including 45 {\it K2}, 1 DBSP, 14 MuSCAT2, 4 TRAPPIST, and 2 Euler, which are presented in Table~\ref{tab:tmid}. We fitted a linear ephemeris to the 66 mid-transit times using \texttt{emcee}, where an error multiple was adopted to rescale the error bars of the mid-transit times. This resulted in a reduced chi-square of $\chi^2_\nu=1.36$ based on the original error bars. The residuals have a standard deviation of 19.5~sec. 

Figure~\ref{fig:ephemeris} presents the mid-transit time residuals assuming the best-fitting linear ephemeris. For comparison, we derived the transit ephemeris based on the {\it K2} data only or the MuSCAT2 data only, respectively. The two individually derived transit ephemerides are well consistent with each other as can be seen in the zoom-in views of Fig.~\ref{fig:ephemeris}. In particular, the MuSCAT2-based ephemeris is rather precise, thanks to a broader time coverage. Therefore, the current observations do not reveal any significant transit timing variations, confirming the conclusion of \citet{2018AJ....156...44M}. 

\subsubsection{Physical parameters}

To revise the physical parameters, we performed a joint analysis of radial velocities (RV), transits, stellar evolution tracks and spectral energy distribution (SED) using the \textsc{idl} package \texttt{EXOFASTv2}\footnote{\url{https://github.com/jdeast/EXOFASTv2}} \citep{2013PASP..125...83E,2019arXiv190709480E}. We collected 21 RV measurements by the CORALIE and SOPHIE spectrographs presented in the discovery paper \citep{2014A&A...570A..64S}. To reduce the computation cost, we only used the light curves of {\it K2}, MuSCAT2 $g$, $r$, $i$, $z_s$, and phase-folded them in 1~min intervals individually, after removing the best-fitting noise models determined in Sect~\ref{sec:lcfit}. We assumed a circular orbit for WASP-104b. We used the MIST stellar evolutionary models \citep{2016ApJS..222....8D}, and adopted the latest spectroscopic stellar parameters of $T_\mathrm{eff}=5416\pm 86$~K and $\mathrm{[Fe/H]}=0.40\pm 0.06$ from \citet{2018A&A...620A..58S} as priors. Other priors were the parallax value of $5.34\pm 0.04$~mas from {\it Gaia} DR2 \citep{2018A&A...616A...1G} and a $V$-band extinction upper limit of 0.09176~mag from \citet{2011ApJ...737..103S}. For the SED fitting, we used magnitudes measured in $B$, $V$, $g$, $r$, $i$ from the URAT1 catalog \citep{2015AJ....150..101Z}, $J$, $H$, $K$ from the 2MASS catalog \citep{2003yCat.2246....0C}, and $W1$, $W2$, $W3$ from the AllWISE catalog \citep{2014yCat.2328....0C}. 

Table~\ref{tab:exofast} presents the physical parameters derived from the posterior distributions sampled by \texttt{EXOFASTv2}. In general, these parameters agree well with those reported by \citet{2018AJ....156...44M}, except for the planet mass, which is probably a typo (i.e., 1.211~M$_\mathrm{Jup}$ instead of 1.311~M$_\mathrm{Jup}$) in \citet{2018AJ....156...44M}.

\begin{table}
\centering
\caption{Derived physical parameters for the WASP-104 system from \texttt{EXOFASTv2}.}
\label{tab:exofast}
\begin{tabular}{p{0.5cm}cr}
\hline\hline
Symbol & Parameter and unit & Value\\
\hline
\multicolumn{3}{l}{Stellar Parameters:}\\\noalign{\smallskip}
~~$M_\star$ & Mass [M$_\odot$] & $1.016\pm 0.033$ \\\noalign{\smallskip}
~~$R_\star$ & Radius [R$_\odot$] & $0.935\pm 0.010$\\\noalign{\smallskip}
~~$L_\star$ & Luminosity [L$_\odot$] & $0.645\pm 0.014$\\\noalign{\smallskip}
~~$\rho_\star$ & Density [cgs] & $1.7515\pm 0.0081$\\\noalign{\smallskip}
~~$\log{g_\star}$ & Surface gravity [cgs] & $4.5032\pm 0.0050$\\\noalign{\smallskip}
~~$T_{\rm eff}$ & Effective temperature [K] & $5348\pm 32$\\\noalign{\smallskip}
~~$[{\rm Fe/H}]$ & Metallicity [dex] &$0.410\pm 0.057$\\\noalign{\smallskip}
~~$Age$ & Age [Gyr] & $2.3^{+2.1}_{-1.5}$\\\noalign{\smallskip}
~~$d$ & Distance [pc] & $186.4\pm1.3$\\\noalign{\smallskip}
~~$K_\star$ & RV semi-amplitude [m\,s$^{-1}$] &$201.4\pm 5.8$\\\noalign{\smallskip}
\multicolumn{3}{l}{Planetary Parameters:}\\\noalign{\smallskip}
~~$M_\mathrm{p}$ & Mass [M$_\mathrm{Jup}$] & $1.216\pm 0.045$\\\noalign{\smallskip}
~~$R_\mathrm{p}$ & Radius [R$_\mathrm{Jup}$] & $1.093\pm 0.012$\\\noalign{\smallskip}
~~$\rho_\mathrm{p}$ & Density [cgs] & $1.158\pm 0.037$\\\noalign{\smallskip}
~~$\log{g_\mathrm{p}}$ & Surface gravity [cgs] & $3.403\pm 0.013$\\\noalign{\smallskip}
~~$g_\mathrm{p}$ & Surface gravity [m\,s$^{-2}$] & $25.26\pm 0.74$\\\noalign{\smallskip}
~~$T_\mathrm{eq}$ & Equilibrium temperature [K] & $1474\pm 9$\\\noalign{\smallskip}
~~$a$ & Semi-major axis [AU] & $0.02864\pm 0.00031$\\\noalign{\smallskip}
~~$\langle F \rangle$ & Incident Flux [10$^9$~erg\,s$^{-1}$\,cm$^{-2}$] & $1.070\pm 0.025$\\\noalign{\smallskip}
\hline
\end{tabular}
\end{table}

\subsection{Transmission spectrum}

The DBSP transmission spectrum does not reveal any significant excess absorption at Na, K, or other atoms or molecules that are expected to be common in hot Jupiter atmospheres. When comparing the MuSCAT2 transmission spectrum to that of DBSP, they show the same overall spectral shape, except for a constant offset of $\Delta R_\mathrm{p}/R_\star=0.00234\pm0.00022$ between them. This offset could be introduced by the rotational modulation of the host star, which requires a flux variability at the $\sim$4 per\,cent level. However, \citet{2018AJ....156...44M} found a rotational modulation with a semi-amplitude of $\sim$0.04 per\,cent from the {\it K2} data. Considering the large time separation, it could also be due to a stellar cycle of 8--10 yr. Another possibility would be due to different systematics, which could potentially bias the measurements. In particular, there is only one DBSP transit observation. 

\subsubsection{Comparison to simplified model assumptions}

Allowing a constant offset correction on the DBSP data, the DBSP and MuSCAT2 combined transmission spectrum of WASP-104b is very similar to that of the puffy hot Jupiter WASP-31b \citep{2017MNRAS.467.4591G}. The latter has a low gravity of 4.56~m\,s$^{-1}$ \citep{2011A&A...531A..60A} and exhibits an enhanced slope \citep{2019ApJ...887L..20W,2020ApJ...895L..47O} at wavelengths shorter than $\sim$530~nm and a cloud deck at longer wavelengths. Following \citet{2017MNRAS.467.4591G}, we consider the model assumptions: ({\it i}) a flat-line model, ({\it ii}) a sloped-line model, and ({\it iii}) a two-component model (i.e., sloped- and flat-lines for the shorter and longer wavelengths, respectively). In all three cases, the DBSP offset is a free parameter. We performed the model fitting using the \textsc{Python} package \texttt{PyMultiNest}\footnote{\url{https://johannesbuchner.github.io/PyMultiNest}} \citep{2014A&A...564A.125B}, which relies on the MultiNest library \citep{2009MNRAS.398.1601F}. This package implements the multimodal nested sampling algorithm, and allows us to calculate the model evidence $\ln\mathcal{Z}$. As shown in Table \ref{tab:model_cmp}, the two-component model is significantly preferred over the flat-line model ($\Delta\ln\mathcal{Z}=9.7$) and the sloped-line model ($\Delta\ln\mathcal{Z}=6.4$).

The observed slope of a transmission spectrum can be linked to the index $\alpha$ \citep[also known as the scattering slope in some studies, e.g.,][]{2019PASP..131c4501Z} through $\mathrm{d}R_\mathrm{p}/\mathrm{d}\ln\lambda=\alpha H$ if it is assumed to originate from the power law scattering cross section $\kappa=\kappa_0(\lambda/\lambda_0)^\alpha$ \citep{2008A&A...481L..83L}, where $H=k_\mathrm{B}T/(\mu g_\mathrm{p})$ is the scale height. Consequently, the index $\alpha$ can be derived as:
\begin{equation}
\alpha=\frac{\mu g_\mathrm{p}R_\star}{k_\mathrm{B}T}\frac{\mathrm{d}R_\mathrm{p}/R_\star}{\mathrm{d}\ln\lambda},
\end{equation}
where $k_\mathrm{B}$ is Boltzmann constant. We adopted $\mu=2.3$~g\,mol$^{-1}$ for the mean molecular weight, and obtained the values of $g_\mathrm{p}$, $T(=T_\mathrm{eq})$, $R_\star$ from Table~\ref{tab:exofast}. As a result, we derived a scattering slope of $\alpha=-7.3\pm 1.9$ over the whole optical wavelength range based on the sloped-line model, and $\alpha=-17.6 ^{+3.7}_{-3.9}$ for the wavelengths shorter than $\sim$630~nm based on the two-component model.

\begin{table}
\centering
\caption{Bayesian model comparison.}
\label{tab:model_cmp}
\begin{tabular}{ccccl}
\hline\hline\noalign{\smallskip}
Model & $\ln\mathcal{Z}$ & $\chi^2_\mathrm{MAP}$ $^a$ & dof & Description\\\noalign{\smallskip}
\hline\noalign{\smallskip}
{\it i}   & 151.161 & 48.9 & 22 & Flat-line model\\\noalign{\smallskip}
\hline\noalign{\smallskip}
{\it ii}  & 154.426 & 33.9 & 21 & Sloped-line model\\\noalign{\smallskip}
\hline\noalign{\smallskip}
{\it iii} & 160.854 & 21.4 & 20 & Two-component model\\\noalign{\smallskip}
\hline\noalign{\smallskip}
{\it iv}  & -- & -- & -- & \texttt{petitRADTRANS}, no spot\\\noalign{\smallskip}
  & 165.907 & 13.2 & 13 & ~~Full model\\\noalign{\smallskip}
  & 165.808 & 16.3 & 14 & ~~No H$_2$O\\\noalign{\smallskip}
  & 166.807 & 13.9 & 14 & ~~No TiO\\\noalign{\smallskip}
  & 166.239 & 13.1 & 14 & ~~No VO\\\noalign{\smallskip}
  & 165.940 & 14.5 & 14 & ~~No Na\\\noalign{\smallskip}
  & 164.591 & 17.0 & 14 & ~~No K\\\noalign{\smallskip}
\hline\noalign{\smallskip}
{\it v}   & 162.004 & 19.7 & 16 & \texttt{PLATON}, no spot\\\noalign{\smallskip}
\hline\noalign{\smallskip}
{\it vi}  & 160.207 & 23.0 & 20 &\texttt{PLATON}, pure spot\\\noalign{\smallskip}
\hline
\end{tabular}
\begin{flushleft}
{\small \textit{Notes.} 
$^{a}$$\chi^2$ for the Maximum a Posteriori (MAP) model.
}
\end{flushleft}
\end{table}

\subsubsection{Retrieval assuming planetary atmosphere}
\label{sec:retrieval_atm}

In the following subsections, we perform spectral retrieval analyses on the DBSP and MuSCAT2 combined transmission spectrum, to explore the hazy atmosphere scenario and the spot contamination scenario. 

We first use the \textsc{Python} package \texttt{PLATON}\footnote{\url{https://github.com/ideasrule/platon}} \citep{2019PASP..131c4501Z,2020ApJ...899...27Z} to perform the spectral retrieval analysis. This package assumes equilibrium chemistry models computed with \texttt{GGchem} \citep{2018A&A...614A...1W}, and allows two free parameters (metallicity $Z$ and C/O ratio) to specify the atmospheric composition. The clouds and hazes are parameterised as a gray absorption at the pressure $P_\mathrm{cloud}$ and a power law cross section (with a scattering slope of $\alpha$ and an enhancement factor $A_\mathrm{scatt}$ over nominal H$_2$ Rayleigh scattering), respectively. In addition, DBSP offset ($\Delta D_\mathrm{DBSP}=\Delta R_\mathrm{p}^2/R_\star^2$), isothermal temperature $T$ and planet radius $R_\mathrm{p}$ at 1~bar are free parameters as well. 

We also use the \textsc{Python} package \texttt{petitRADTRANS}\footnote{\url{https://gitlab.com/mauricemolli/petitRADTRANS}} \citep{2019A&A...627A..67M} to perform another run of spectral retrieval analysis. This package allows us to specify atmospheric composition in terms of individual atoms and molecules. We use a similar clouds and hazes parameterisation, except that the scattering amplitude is expressed in the form of opacity in units of cm$^2$\,g$^{-1}$. The included opacity sources are H$_2$O, TiO, VO, Na, K, and their mass fractions are free parameters. The remaining free parameters are: isothermal temperature $T$, reference pressure $P_0$ (corresponding to white-light radius), and DBSP offset $\Delta D_\mathrm{DBSP}$.

\begin{figure*}
\centering
\includegraphics[width=0.9\textwidth]{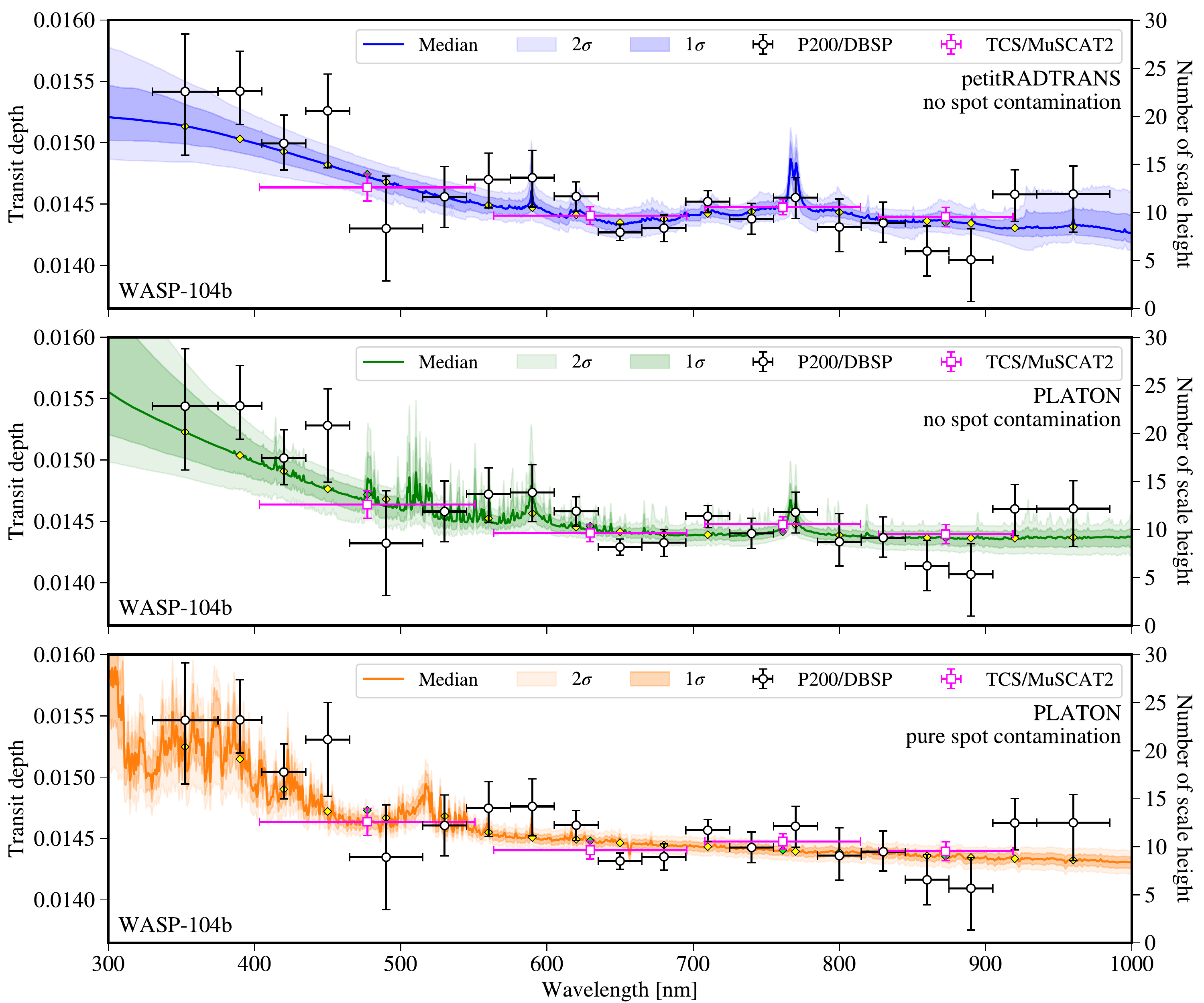}
\caption{Transmission spectrum of the hot Jupiter WASP-104b. The transit depths measured by P200/DBSP and TCS/MuSCAT2 are shown as black circles and magenta squares, respectively. {\it Top panel} presents the retrieval analysis performed using the \texttt{petitRADTRANS} package \citep{2019A&A...627A..67M}, where the planetary atmosphere is assumed to induce any spectral features in the measurements. {\it Middle} and {\it bottom} panels present the retrieval analyses performed using the \texttt{PLATON} package \citep{2019PASP..131c4501Z,2020ApJ...899...27Z}, where the contribution of spot contamination is excluded ({\it middle panel}) or is the only cause of transit-depth variation ({\it bottom panel}). For all three panels, the DBSP spectrum has been corrected for the overall offset determined in each retrieval analysis.}
\label{fig:retrieval}
\end{figure*}

For both retrieval runs, we employ \texttt{PyMultiNest} to perform the fitting processes. The resulting statistics and parameters are listed in Tables \ref{tab:model_cmp} and \ref{tab:retrieved_params}. The retrieved models and confidence regions are shown in Fig.~\ref{fig:retrieval}, and corresponding posterior distributions are given in Fig.~\ref{fig:corner_petit} and Fig.~\ref{fig:corner_platon_nospot}. 

The two retrieval runs resulted in consistent results. No atoms or molecules can be detected under current data precision, although the data disfavour TiO but favour K tentatively at $\sim$2$\sigma$ level. Similarly, the C/O ratio and metallicity cannot be constrained given the absence of significant spectral features. This indicates that it is very challenging to detect the atmospheres of planets with intermediate surface gravity ($\sim$25~m\,s$^{-2}$ in the case of WASP-104b), because their atmospheres are relatively compact and will require much higher data precision to resolve spectral features.

\begin{table}
\centering
\caption{Parameters derived from spectral retrieval analyses.}
\label{tab:retrieved_params}
\begin{tabular}{cccc}
\hline\hline
Parameters & \texttt{petitRADTRANS} & \multicolumn{2}{c}{\texttt{PLATON}} \\\noalign{\smallskip}
\cline{3-4}\noalign{\smallskip}
& no spot & no spot & pure spot\\
\hline\noalign{\smallskip}
$T$ [K]                          & $1966 ^{+479}_{-422}$ & $1969 ^{+604}_{-623}$ & -- \\\noalign{\smallskip}
$R_\mathrm{p}$ [$R_\mathrm{J}$]  & -- & $1.078 ^{+0.006}_{-0.012}$ & $1.098 ^{+0.005}_{-0.005}$ \\\noalign{\smallskip}
$\log P_0 \mathrm{[bar]}$        & $-0.06 ^{+0.95}_{-0.83}$ & -- & -- \\\noalign{\smallskip}
$\log P_\mathrm{cloud} \mathrm{[bar]}$  & $-0.55 ^{+1.6}_{-1.7}$ & $-1.3 ^{+2.1}_{-1.5}$ & -- \\\noalign{\smallskip}
$\alpha$                         & $-16.6 ^{+3.1}_{-2.3}$ & $-17.2 ^{+3.3}_{-1.9}$ & -- \\\noalign{\smallskip}
$\log A_\mathrm{scatt}$$^a$      & $1.63 ^{+0.83}_{-0.88}$ & $-2.0 ^{+1.5}_{-1.2}$ & -- \\\noalign{\smallskip}
$\mathrm{C/O}$                   & -- & $1.38 ^{+0.41}_{-0.68}$ & -- \\\noalign{\smallskip}
$\log Z/Z_\odot$                 & -- & $0.27 ^{+1.14}_{-0.88}$ & -- \\\noalign{\smallskip}
$\log X_{\mathrm{H}_2\mathrm{O}}$ & $-4.3 ^{+2.7}_{-3.6}$ & -- & -- \\\noalign{\smallskip}
$\log X_\mathrm{TiO}$            & $-8.4 ^{+1.0}_{-1.0}$ & -- & -- \\\noalign{\smallskip}
$\log X_\mathrm{VO}$             & $-8.3 ^{+1.3}_{-1.1}$ & -- & -- \\\noalign{\smallskip}
$\log X_\mathrm{Na}$             & $-5.7 ^{+2.7}_{-2.7}$ & -- & -- \\\noalign{\smallskip}
$\log X_\mathrm{K}$              & $-3.1 ^{+1.5}_{-2.1}$ & -- & -- \\\noalign{\smallskip}
$T_\mathrm{spot} \mathrm{[K]}$   & -- & -- & $5163 ^{+79}_{-191}$ \\\noalign{\smallskip}
$f_\mathrm{spot}$                & -- & -- & $0.26 ^{+0.17}_{-0.11}$ \\\noalign{\smallskip}
$\Delta D_\mathrm{DBSP}$ [ppm]   & $-539 ^{+48}_{-48}$ & $-560 ^{+48}_{-48}$ & $-586 ^{+51}_{-52}$ \\\noalign{\smallskip}
\hline
\end{tabular}
\begin{flushleft}
{\small\color{black} \textit{Notes.} 
$^{a}$For \texttt{petitRADTRANS}, $f_\mathrm{scatt}=\kappa_0$ is a physical quantity. For \texttt{PLATON}, $f_\mathrm{scatt}$ is a dimensionless enhancement factor over the nominal H$_2$ Rayleigh scattering.
}
\end{flushleft}
\end{table}

\subsubsection{Retrieval assuming stellar heterogeneity}

We also perform a third spectral retrieval analysis with \texttt{PLATON}, where the planet is assumed to have no atmosphere and the spot contamination is fully responsible for the variation in wavelength dependent transit depths. The adopted free parameters are planet radius $R_\mathrm{p}$, DBSP offset $\Delta D_\mathrm{DBSP}$, spot temperature $T_\mathrm{spot}$, and spot coverage $f_\mathrm{spot}$. The spot contamination is implemented by \texttt{PLATON} using
\begin{equation}
D_{\lambda,c} = D_\lambda\frac{S(\lambda,T_\mathrm{phot})}{f_\mathrm{spot}S(\lambda,T_\mathrm{spot})+(1-f_\mathrm{spot})S(\lambda,T_\mathrm{phot})},
\end{equation}
where $D_\lambda$ is the uncontaminated true transit depth at the wavelength $\lambda$, $T_\mathrm{phot}=5348$~K is the adopted temperature for unspotted stellar photosphere, and $S(\lambda,T)$ are the interpolated BT-NextGen (AGSS2009) stellar models \citep{2012RSPTA.370.2765A}.

The retrieved models are shown in the third row of Fig.~\ref{fig:retrieval}. The retrieved parameters are given in Table \ref{tab:retrieved_params}, and their posterior distributions in Fig.~\ref{fig:corner_platon_spot}. The best retrieved models indicate that the spot temperature is $185 ^{+79}_{-191}$~K lower than the photosphere and that the spot coverage is $26 ^{+17}_{-11}$ per\,cent. The retrieved spot temperature does not follow the trend observed for active giants and dwarfs that hotter photospheres have larger spot-photosphere temperature contrasts \citep{2005LRSP....2....8B}. Should it follow the trend, with a photosphere temperature of 5348~K, the spot temperature would be $\sim$1500~K lower than the photosphere.

\subsubsection{Interpretation of the slope}

WASP-104b is one of the few transiting planets with intermediate gravity for which a transmission spectrum has been measured. Although the regime of intermediate gravity is barely studied, the current sample tends to show super steep scattering slopes, such as HD 189733b \citep[21.5~m\,s$^{-2}$;][]{2011MNRAS.416.1443S,2013MNRAS.432.2917P}, TrES-3b \citep[27.4~m\,s$^{-2}$;][]{2016A&A...585A.114P}, and WASP-43b \citep[47~m\,s$^{-2}$;][]{2020AJ....159...13W}. Super steep scattering slopes have also been observed in some hot Jupiters with low gravity. For example, \citet{2018AJ....156..283E} found that WASP-121b (9.4~m\,s$^{-2}$) shows a steep slope at $\lambda\leq470$~nm and suggested SH as the absorber. \citet{2020AJ....159....7M} observed extremely large and variable slopes ($\alpha\sim-100,-40$) in two transits of HATS-8b (4.5~m\,s$^{-2}$) which are difficult to explain. For several of the planets with super steep scattering slopes, the debate is open on whether potential spot contamination induced by their active host stars can be underlying responsible physical mechanism, including HD 189733b \citep{2014ApJ...791...55M}, WASP-43b \citep{2020AJ....159...13W}, and WASP-19b \citep{2019MNRAS.482.2065E}. 

In our spectral retrieval analyses, both spot contamination and hazy atmosphere models can fit WASP-104b's transmission spectrum reasonably well, although the Bayesian model comparison (Table \ref{tab:model_cmp}) favours the hazy atmosphere models. On the other hand, WASP-104 does not show any signs of significant stellar activity \citep{2014A&A...570A..64S,2018AJ....156...44M}, only exhibiting a rotational modulation with a semi-amplitude of $\sim$400~ppm. Should the slope be induced by the spots, with a derived coverage of $26 ^{+17}_{-11}$ per\,cent, the spots have to be small and uniformly distributed or contain a large fraction of non-modulating component to avoid inducing large flux variability \citep{2018ApJ...853..122R,2019AJ....157...96R,2020A&A...641A..82R}. This is consistent with the fact that none of our observed transit light curves have revealed any significant spot-crossing events, nor have the 45 {\it K2} transit light curves of higher precision. With a large set of observations dispersed in three years, the MuSCAT2 multi-colour transit depths should have averaged out the variation of spot contamination. Since the MuSCAT2 spectral shape is consistent with that of DBSP, spots of similar properties would have to be persistent for at least five years. The large spot coverage would indicate a young age for the host star \citep[e.g.,][]{2020ApJ...893...67M}, which is not favoured by the low lithium abundance \citep{2014A&A...570A..64S} and the slow rotation period \citep[$\sim$23 or $\sim$46 days;][]{2018AJ....156...44M}. However, although the full set of evidences point out that the observed slope is not likely caused by spot contamination, we suggest to acquire further evidence, through follow-up transmission spectroscopy and detailed characterisation of the host star's activity, to completely rule out its possibility.

While in theory spot contamination could result in a wide range of slope values,  planetary atmospheres can exhibit steep slope values only under certain conditions. \citet{2017MNRAS.471.4355P} suggest that a small number of sulphide condensates of modal sizes $\sim$10$^{-2}$~$\mu$m could introduce very steep slopes. For example, MnS could result in a slope as steep as $-$13. But a recent study shows that metal sulfide clouds are strongly inhibited due to nucleation energy barriers \citep{2020NatAs.tmp..114G}. On the other hand, photochemical haze can also potentially steepen the scattering slope if eddy diffusion is efficient and the haze mass flux is moderate \citep{2019ApJ...877..109K,2020ApJ...895L..47O}. Planets with equilibrium temperatures of 1000-1500~K are suggested to be potential targets for super-Rayleigh slopes generated by photochemical haze. With an equilibrium temperature of $1474 \pm 9$~K, WASP-104b falls into this temperature range, and would be in a location very close to WASP-31b in the figure 5 of \citet{2020ApJ...895L..47O}. According to \citet{2020ApJ...895L..47O}, the observed slope of WASP-104b could be introduced by tholin at a haze mass flux of $F\sim 10^{-12}$~g\,cm$^2$\,s$^{-1}$ and an eddy diffusion of $K_z\sim10^{-9}$~cm$^2$\,s$^{-1}$.

\section{Conclusions}
\label{sec:conclusion}

We acquired 51 broadband light curves of WASP-104b by observing one transit of WASP-104b with the blue and red channels of the low-resolution spectrograph DBSP at the Palomar 200-inch telescope, and 14 transits with the multi-colour imager MuSCAT2 at the 1.52~m Telescopio Carlos S\'{a}nchez. We derived the flux contamination by the companion star close to WASP-104 using the DBSP spectra. We re-analysed the 45 {\it K2} transit light curves after considering the dilution correction. Based on the collection of DBSP, MuSCAT2, and {\it K2} light curves, we were able to update the transit parameters and physical parameters. We confirm that the planet does not exhibit any transit timing variations.

We further divided the DBSP spectra into chromatic passbands. The resulting DBSP transmission spectrum reveals an enhanced slope at wavelengths shorter than $\sim$630~nm and a relatively flat spectrum at longer wavelengths. The broadband transmission spectrum acquired by MuSCAT2 is consistent with that of DBSP in the overall spectral shape except for a constant offset. We performed spectral retrieval analyses on the DBSP and MuSCAT2 combined transmission spectrum using two different publicly available codes, and the results were in a broad agreement. To find the underlying physical cause of the measured slope, we explored the scenarios where the slope is caused by the planet's atmosphere or by spot contamination. We find that stellar spot contamination is a less likely explanation given the knowledge about the host star's activity and the statistics from the Bayesian spectral retrieval analyses. However, further evidence is required to completely rule out the possibility of spot contamination. 

Hot Jupiters with relatively high gravity, like WASP-104b, are rarely studied in transmission spectroscopy. Thus their observational signatures of gravity on clouds/hazes are still not well studied, especially in the upper atmospheres. A combination of better ground-based transit measurements, such as the one performed here, and the extraordinary precision of the {\it James Webb} space telescope in the coming years, will hopefully allow us to explore and understand this population.

\section*{Acknowledgements}

    G.\,C. acknowledges the support by the B-type Strategic Priority Program of the Chinese Academy of Sciences (Grant No. XDB41000000), the National Natural Science Foundation of China (Grant No. 42075122), the Natural Science Foundation of Jiangsu Province (Grant No. BK20190110), and the Minor Planet Foundation of the Purple Mountain Observatory. 
    This work is partly financed by the Spanish Ministry of Economics and Competitiveness through grants PGC2018-098153-B-C31 and PID2019-109522GB-53.
    This work is partly supported by JSPS KAKENHI Grant Numbers JP18H01265, JP18H05439, JP17H04574, and JST PRESTO Grant Number JPMJPR1775.
    M.T. is supported by JSPS KAKENHI Grant Numbers 18H05442, 15H02063, and 22000005. 
    This work is partly supported by Grant-in-Aid for JSPS Fellows, Grant Number JP20J21872.
    Z.G. was supported by the VEGA grant of the Slovak Academy of Sciences No. 2/0031/18, by the Hungarian NKFI grant No. K-119517 and 
the GINOP No. 2.3.2-15-2016-00003 of the Hungarian National Research, Development and Innovation Office, and by the City of Szombathely under agreement No. 67.177-21/2016.
    The authors thank the anonymous referee for constructive comments on the manuscript.

    This work uses data obtained through the Telescope Access Program (TAP), which has been funded by the National Astronomical Observatories, Chinese Academy of Sciences, and the Special Fund for Astronomy from the Ministry of Finance.
    Observations obtained with the Hale Telescope at Palomar Observatory were obtained as part of an agreement between the National Astronomical Observatories, Chinese Academy of Sciences, and the California Institute of Technology.
    This work is partly based on observations made with the MuSCAT2 instrument, developed by ABC, at Telescopio Carlos S\'{a}nchez operated on the island of Tenerife by the IAC in the Spanish Observatorio del Teide. 
    This work includes data collected by the Kepler mission and obtained from the MAST data archive at the Space Telescope Science Institute (STScI). Funding for the Kepler mission is provided by the NASA Science Mission Directorate. STScI is operated by the Association of Universities for Research in Astronomy, Inc., under NASA contract NAS 5-26555.
    This work has made use of data from the European Space Agency (ESA) mission {\it Gaia} (\url{https://www.cosmos.esa.int/gaia}), processed by the {\it Gaia} Data Processing and Analysis Consortium (DPAC, \url{https://www.cosmos.esa.int/web/gaia/dpac/consortium}). Funding for the DPAC has been provided by national institutions, in particular the institutions participating in the {\it Gaia} Multilateral Agreement.
    This work has made use of Matplotlib \citep{2007CSE.....9...90H}, the VizieR catalog access tool, CDS, Strasbourg, France \citep{2000A&AS..143...23O}, and TEPCat \citep{2011MNRAS.417.2166S}. 
    
\section*{Data Availability}
The data underlying this article will be shared on reasonable request to the corresponding author. The reduced light curves presented in this work will be made available at the CDS (\url{http://cdsarc.u-strasbg.fr/}).




\bibliographystyle{mnras}
\bibliography{ref_db.bib} 

\begin{thebibliography}{}
\makeatletter
\relax
\def\mn@urlcharsother{\let\do\@makeother \do\$\do\&\do\#\do\^\do\_\do\%\do\~}
\def\mn@doi{\begingroup\mn@urlcharsother \@ifnextchar [ {\mn@doi@}
  {\mn@doi@[]}}
\def\mn@doi@[#1]#2{\def\@tempa{#1}\ifx\@tempa\@empty \href
  {http://dx.doi.org/#2} {doi:#2}\else \href {http://dx.doi.org/#2} {#1}\fi
  \endgroup}
\def\mn@eprint#1#2{\mn@eprint@#1:#2::\@nil}
\def\mn@eprint@arXiv#1{\href {http://arxiv.org/abs/#1} {{\tt arXiv:#1}}}
\def\mn@eprint@dblp#1{\href {http://dblp.uni-trier.de/rec/bibtex/#1.xml}
  {dblp:#1}}
\def\mn@eprint@#1:#2:#3:#4\@nil{\def\@tempa {#1}\def\@tempb {#2}\def\@tempc
  {#3}\ifx \@tempc \@empty \let \@tempc \@tempb \let \@tempb \@tempa \fi \ifx
  \@tempb \@empty \def\@tempb {arXiv}\fi \@ifundefined
  {mn@eprint@\@tempb}{\@tempb:\@tempc}{\expandafter \expandafter \csname
  mn@eprint@\@tempb\endcsname \expandafter{\@tempc}}}

\bibitem[\protect\citeauthoryear{{Akaike}}{{Akaike}}{1974}]{1974ITAC...19..716A}
{Akaike} H.,  1974, IEEE Transactions on Automatic Control, \href
  {https://ui.adsabs.harvard.edu/abs/1974ITAC...19..716A} {19, 716}

\bibitem[\protect\citeauthoryear{{Alam} et~al.,}{{Alam}
  et~al.}{2020}]{2020AJ....160...51A}
{Alam} M.~K.,  et~al., 2020, \mn@doi [\aj] {10.3847/1538-3881/ab96cb}, \href
  {https://ui.adsabs.harvard.edu/abs/2020AJ....160...51A} {160, 51}

\bibitem[\protect\citeauthoryear{{Alexoudi} et~al.,}{{Alexoudi}
  et~al.}{2018}]{2018A&A...620A.142A}
{Alexoudi} X.,  et~al., 2018, \mn@doi [\aap] {10.1051/0004-6361/201833691},
  \href {https://ui.adsabs.harvard.edu/abs/2018A&A...620A.142A} {620, A142}

\bibitem[\protect\citeauthoryear{{Allard}, {Homeier}  \& {Freytag}}{{Allard}
  et~al.}{2012}]{2012RSPTA.370.2765A}
{Allard} F.,  {Homeier} D.,   {Freytag} B.,  2012, \mn@doi [Philosophical
  Transactions of the Royal Society of London Series A]
  {10.1098/rsta.2011.0269}, \href
  {https://ui.adsabs.harvard.edu/abs/2012RSPTA.370.2765A} {370, 2765}

\bibitem[\protect\citeauthoryear{{Ambikasaran}, {Foreman-Mackey}, {Greengard},
  {Hogg}  \& {O'Neil}}{{Ambikasaran} et~al.}{2015}]{2015ITPAM..38..252A}
{Ambikasaran} S.,  {Foreman-Mackey} D.,  {Greengard} L.,  {Hogg} D.~W.,
  {O'Neil} M.,  2015, \mn@doi [IEEE Transactions on Pattern Analysis and
  Machine Intelligence] {10.1109/TPAMI.2015.2448083}, \href
  {https://ui.adsabs.harvard.edu/abs/2015ITPAM..38..252A} {38, 252}

\bibitem[\protect\citeauthoryear{{Anderson} et~al.,}{{Anderson}
  et~al.}{2011}]{2011A&A...531A..60A}
{Anderson} D.~R.,  et~al., 2011, \mn@doi [\aap] {10.1051/0004-6361/201016208},
  \href {https://ui.adsabs.harvard.edu/abs/2011A&A...531A..60A} {531, A60}

\bibitem[\protect\citeauthoryear{{Berdyugina}}{{Berdyugina}}{2005}]{2005LRSP....2....8B}
{Berdyugina} S.~V.,  2005, \mn@doi [Living Reviews in Solar Physics]
  {10.12942/lrsp-2005-8}, \href
  {https://ui.adsabs.harvard.edu/abs/2005LRSP....2....8B} {2, 8}

\bibitem[\protect\citeauthoryear{{Buchner} et~al.,}{{Buchner}
  et~al.}{2014}]{2014A&A...564A.125B}
{Buchner} J.,  et~al., 2014, \mn@doi [\aap] {10.1051/0004-6361/201322971},
  \href {https://ui.adsabs.harvard.edu/abs/2014A&A...564A.125B} {564, A125}

\bibitem[\protect\citeauthoryear{{Carter} et~al.,}{{Carter}
  et~al.}{2020}]{2020MNRAS.494.5449C}
{Carter} A.~L.,  et~al., 2020, \mn@doi [\mnras] {10.1093/mnras/staa1078}, \href
  {https://ui.adsabs.harvard.edu/abs/2020MNRAS.494.5449C} {494, 5449}

\bibitem[\protect\citeauthoryear{{Charbonneau}, {Brown}, {Noyes}  \&
  {Gilliland}}{{Charbonneau} et~al.}{2002}]{2002ApJ...568..377C}
{Charbonneau} D.,  {Brown} T.~M.,  {Noyes} R.~W.,   {Gilliland} R.~L.,  2002,
  \mn@doi [\apj] {10.1086/338770}, \href
  {https://ui.adsabs.harvard.edu/abs/2002ApJ...568..377C} {568, 377}

\bibitem[\protect\citeauthoryear{{Chen} et~al.,}{{Chen}
  et~al.}{2014}]{2014A&A...563A..40C}
{Chen} G.,  et~al., 2014, \mn@doi [\aap] {10.1051/0004-6361/201322740}, \href
  {https://ui.adsabs.harvard.edu/abs/2014A&A...563A..40C} {563, A40}

\bibitem[\protect\citeauthoryear{{Chen}, {Pall{\'e}}, {Nortmann}, {Murgas},
  {Parviainen}  \& {Nowak}}{{Chen} et~al.}{2017a}]{2017A&A...600L..11C}
{Chen} G.,  {Pall{\'e}} E.,  {Nortmann} L.,  {Murgas} F.,  {Parviainen} H.,
  {Nowak} G.,  2017a, \mn@doi [\aap] {10.1051/0004-6361/201730736}, \href
  {https://ui.adsabs.harvard.edu/abs/2017A&A...600L..11C} {600, L11}

\bibitem[\protect\citeauthoryear{{Chen}, {Guenther}, {Pall{\'e}}, {Nortmann},
  {Nowak}, {Kunz}, {Parviainen}  \& {Murgas}}{{Chen}
  et~al.}{2017b}]{2017A&A...600A.138C}
{Chen} G.,  {Guenther} E.~W.,  {Pall{\'e}} E.,  {Nortmann} L.,  {Nowak} G.,
  {Kunz} S.,  {Parviainen} H.,   {Murgas} F.,  2017b, \mn@doi [\aap]
  {10.1051/0004-6361/201630228}, \href
  {https://ui.adsabs.harvard.edu/abs/2017A&A...600A.138C} {600, A138}

\bibitem[\protect\citeauthoryear{{Chen} et~al.,}{{Chen}
  et~al.}{2018}]{2018A&A...616A.145C}
{Chen} G.,  et~al., 2018, \mn@doi [\aap] {10.1051/0004-6361/201833033}, \href
  {https://ui.adsabs.harvard.edu/abs/2018A&A...616A.145C} {616, A145}

\bibitem[\protect\citeauthoryear{{Chen}, {Casasayas-Barris}, {Pall{\'e}},
  {Welbanks}, {Madhusudhan}, {Luque}  \& {Murgas}}{{Chen}
  et~al.}{2020}]{2020A&A...642A..54C}
{Chen} G.,  {Casasayas-Barris} N.,  {Pall{\'e}} E.,  {Welbanks} L.,
  {Madhusudhan} N.,  {Luque} R.,   {Murgas} F.,  2020, \mn@doi [\aap]
  {10.1051/0004-6361/202038661}, \href
  {https://ui.adsabs.harvard.edu/abs/2020A&A...642A..54C} {642, A54}

\bibitem[\protect\citeauthoryear{{Crossfield} \& {Kreidberg}}{{Crossfield} \&
  {Kreidberg}}{2017}]{2017AJ....154..261C}
{Crossfield} I. J.~M.,  {Kreidberg} L.,  2017, \mn@doi [\aj]
  {10.3847/1538-3881/aa9279}, \href
  {https://ui.adsabs.harvard.edu/abs/2017AJ....154..261C} {154, 261}

\bibitem[\protect\citeauthoryear{{Cutri} \& {et al.}}{{Cutri} \& {et
  al.}}{2014}]{2014yCat.2328....0C}
{Cutri} R.~M.,  {et al.} 2014, VizieR Online Data Catalog, \href
  {https://ui.adsabs.harvard.edu/abs/2014yCat.2328....0C} {p. II/328}

\bibitem[\protect\citeauthoryear{{Cutri} et~al.,}{{Cutri}
  et~al.}{2003}]{2003yCat.2246....0C}
{Cutri} R.~M.,  et~al., 2003, VizieR Online Data Catalog, \href
  {https://ui.adsabs.harvard.edu/abs/2003yCat.2246....0C} {p. II/246}

\bibitem[\protect\citeauthoryear{{Deming} et~al.,}{{Deming}
  et~al.}{2015}]{2015ApJ...805..132D}
{Deming} D.,  et~al., 2015, \mn@doi [\apj] {10.1088/0004-637X/805/2/132}, \href
  {https://ui.adsabs.harvard.edu/abs/2015ApJ...805..132D} {805, 132}

\bibitem[\protect\citeauthoryear{{Dotter}}{{Dotter}}{2016}]{2016ApJS..222....8D}
{Dotter} A.,  2016, \mn@doi [\apjs] {10.3847/0067-0049/222/1/8}, \href
  {https://ui.adsabs.harvard.edu/abs/2016ApJS..222....8D} {222, 8}

\bibitem[\protect\citeauthoryear{{Eastman}, {Siverd}  \& {Gaudi}}{{Eastman}
  et~al.}{2010}]{2010PASP..122..935E}
{Eastman} J.,  {Siverd} R.,   {Gaudi} B.~S.,  2010, \mn@doi [\pasp]
  {10.1086/655938}, \href
  {https://ui.adsabs.harvard.edu/abs/2010PASP..122..935E} {122, 935}

\bibitem[\protect\citeauthoryear{{Eastman}, {Gaudi}  \& {Agol}}{{Eastman}
  et~al.}{2013}]{2013PASP..125...83E}
{Eastman} J.,  {Gaudi} B.~S.,   {Agol} E.,  2013, \mn@doi [\pasp]
  {10.1086/669497}, \href
  {https://ui.adsabs.harvard.edu/abs/2013PASP..125...83E} {125, 83}

\bibitem[\protect\citeauthoryear{{Eastman} et~al.,}{{Eastman}
  et~al.}{2019}]{2019arXiv190709480E}
{Eastman} J.~D.,  et~al., 2019, arXiv e-prints, \href
  {https://ui.adsabs.harvard.edu/abs/2019arXiv190709480E} {p. arXiv:1907.09480}

\bibitem[\protect\citeauthoryear{{Espinoza} \& {Jord{\'a}n}}{{Espinoza} \&
  {Jord{\'a}n}}{2015}]{2015MNRAS.450.1879E}
{Espinoza} N.,  {Jord{\'a}n} A.,  2015, \mn@doi [\mnras]
  {10.1093/mnras/stv744}, \href
  {https://ui.adsabs.harvard.edu/abs/2015MNRAS.450.1879E} {450, 1879}

\bibitem[\protect\citeauthoryear{{Espinoza} et~al.,}{{Espinoza}
  et~al.}{2019}]{2019MNRAS.482.2065E}
{Espinoza} N.,  et~al., 2019, \mn@doi [\mnras] {10.1093/mnras/sty2691}, \href
  {https://ui.adsabs.harvard.edu/abs/2019MNRAS.482.2065E} {482, 2065}

\bibitem[\protect\citeauthoryear{{Evans} et~al.,}{{Evans}
  et~al.}{2018}]{2018AJ....156..283E}
{Evans} T.~M.,  et~al., 2018, \mn@doi [\aj] {10.3847/1538-3881/aaebff}, \href
  {https://ui.adsabs.harvard.edu/abs/2018AJ....156..283E} {156, 283}

\bibitem[\protect\citeauthoryear{{Feroz}, {Hobson}  \& {Bridges}}{{Feroz}
  et~al.}{2009}]{2009MNRAS.398.1601F}
{Feroz} F.,  {Hobson} M.~P.,   {Bridges} M.,  2009, \mn@doi [\mnras]
  {10.1111/j.1365-2966.2009.14548.x}, \href
  {https://ui.adsabs.harvard.edu/abs/2009MNRAS.398.1601F} {398, 1601}

\bibitem[\protect\citeauthoryear{{Foreman-Mackey}, {Hogg}, {Lang}  \&
  {Goodman}}{{Foreman-Mackey} et~al.}{2013}]{2013PASP..125..306F}
{Foreman-Mackey} D.,  {Hogg} D.~W.,  {Lang} D.,   {Goodman} J.,  2013, \mn@doi
  [\pasp] {10.1086/670067}, \href
  {https://ui.adsabs.harvard.edu/abs/2013PASP..125..306F} {125, 306}

\bibitem[\protect\citeauthoryear{{Fortney}}{{Fortney}}{2005}]{2005MNRAS.364..649F}
{Fortney} J.~J.,  2005, \mn@doi [\mnras] {10.1111/j.1365-2966.2005.09587.x},
  \href {https://ui.adsabs.harvard.edu/abs/2005MNRAS.364..649F} {364, 649}

\bibitem[\protect\citeauthoryear{{Fu}, {Deming}, {Knutson}, {Madhusudhan},
  {Mandell}  \& {Fraine}}{{Fu} et~al.}{2017}]{2017ApJ...847L..22F}
{Fu} G.,  {Deming} D.,  {Knutson} H.,  {Madhusudhan} N.,  {Mandell} A.,
  {Fraine} J.,  2017, \mn@doi [\apjl] {10.3847/2041-8213/aa8e40}, \href
  {https://ui.adsabs.harvard.edu/abs/2017ApJ...847L..22F} {847, L22}

\bibitem[\protect\citeauthoryear{{Gaia Collaboration} et~al.,}{{Gaia
  Collaboration} et~al.}{2018}]{2018A&A...616A...1G}
{Gaia Collaboration} et~al., 2018, \mn@doi [\aap]
  {10.1051/0004-6361/201833051}, \href
  {https://ui.adsabs.harvard.edu/abs/2018A&A...616A...1G} {616, A1}

\bibitem[\protect\citeauthoryear{{Gao} et~al.,}{{Gao}
  et~al.}{2020}]{2020NatAs.tmp..114G}
{Gao} P.,  et~al., 2020, \mn@doi [Nature Astronomy]
  {10.1038/s41550-020-1114-3}, \href
  {https://ui.adsabs.harvard.edu/abs/2020NatAs.tmp..114G} {}

\bibitem[\protect\citeauthoryear{{Garhart} et~al.,}{{Garhart}
  et~al.}{2020}]{2020AJ....159..137G}
{Garhart} E.,  et~al., 2020, \mn@doi [\aj] {10.3847/1538-3881/ab6cff}, \href
  {https://ui.adsabs.harvard.edu/abs/2020AJ....159..137G} {159, 137}

\bibitem[\protect\citeauthoryear{{Gibson}, {Aigrain}, {Roberts}, {Evans},
  {Osborne}  \& {Pont}}{{Gibson} et~al.}{2012}]{2012MNRAS.419.2683G}
{Gibson} N.~P.,  {Aigrain} S.,  {Roberts} S.,  {Evans} T.~M.,  {Osborne} M.,
  {Pont} F.,  2012, \mn@doi [\mnras] {10.1111/j.1365-2966.2011.19915.x}, \href
  {https://ui.adsabs.harvard.edu/abs/2012MNRAS.419.2683G} {419, 2683}

\bibitem[\protect\citeauthoryear{{Gibson}, {Aigrain}, {Barstow}, {Evans},
  {Fletcher}  \& {Irwin}}{{Gibson} et~al.}{2013}]{2013MNRAS.428.3680G}
{Gibson} N.~P.,  {Aigrain} S.,  {Barstow} J.~K.,  {Evans} T.~M.,  {Fletcher}
  L.~N.,   {Irwin} P.~G.~J.,  2013, \mn@doi [\mnras] {10.1093/mnras/sts307},
  \href {https://ui.adsabs.harvard.edu/abs/2013MNRAS.428.3680G} {428, 3680}

\bibitem[\protect\citeauthoryear{{Gibson}, {Nikolov}, {Sing}, {Barstow},
  {Evans}, {Kataria}  \& {Wilson}}{{Gibson} et~al.}{2017}]{2017MNRAS.467.4591G}
{Gibson} N.~P.,  {Nikolov} N.,  {Sing} D.~K.,  {Barstow} J.~K.,  {Evans} T.~M.,
   {Kataria} T.,   {Wilson} P.~A.,  2017, \mn@doi [\mnras]
  {10.1093/mnras/stx353}, \href
  {https://ui.adsabs.harvard.edu/abs/2017MNRAS.467.4591G} {467, 4591}

\bibitem[\protect\citeauthoryear{{Greiner} et~al.,}{{Greiner}
  et~al.}{2008}]{2008PASP..120..405G}
{Greiner} J.,  et~al., 2008, \mn@doi [\pasp] {10.1086/587032}, \href
  {https://ui.adsabs.harvard.edu/abs/2008PASP..120..405G} {120, 405}

\bibitem[\protect\citeauthoryear{{Heng}}{{Heng}}{2016}]{2016ApJ...826L..16H}
{Heng} K.,  2016, \mn@doi [\apjl] {10.3847/2041-8205/826/1/L16}, \href
  {https://ui.adsabs.harvard.edu/abs/2016ApJ...826L..16H} {826, L16}

\bibitem[\protect\citeauthoryear{{Horne}}{{Horne}}{1986}]{1986PASP...98..609H}
{Horne} K.,  1986, \mn@doi [\pasp] {10.1086/131801}, \href
  {https://ui.adsabs.harvard.edu/abs/1986PASP...98..609H} {98, 609}

\bibitem[\protect\citeauthoryear{{Howell} et~al.,}{{Howell}
  et~al.}{2014}]{2014PASP..126..398H}
{Howell} S.~B.,  et~al., 2014, \mn@doi [\pasp] {10.1086/676406}, \href
  {https://ui.adsabs.harvard.edu/abs/2014PASP..126..398H} {126, 398}

\bibitem[\protect\citeauthoryear{{Hunter}}{{Hunter}}{2007}]{2007CSE.....9...90H}
{Hunter} J.~D.,  2007, \mn@doi [Computing in Science and Engineering]
  {10.1109/MCSE.2007.55}, \href
  {https://ui.adsabs.harvard.edu/abs/2007CSE.....9...90H} {9, 90}

\bibitem[\protect\citeauthoryear{{Husser}, {Wende-von Berg}, {Dreizler},
  {Homeier}, {Reiners}, {Barman}  \& {Hauschildt}}{{Husser}
  et~al.}{2013}]{2013A&A...553A...6H}
{Husser} T.~O.,  {Wende-von Berg} S.,  {Dreizler} S.,  {Homeier} D.,  {Reiners}
  A.,  {Barman} T.,   {Hauschildt} P.~H.,  2013, \mn@doi [\aap]
  {10.1051/0004-6361/201219058}, \href
  {https://ui.adsabs.harvard.edu/abs/2013A&A...553A...6H} {553, A6}

\bibitem[\protect\citeauthoryear{{Iyer}, {Swain}, {Zellem}, {Line}, {Roudier},
  {Rocha}  \& {Livingston}}{{Iyer} et~al.}{2016}]{2016ApJ...823..109I}
{Iyer} A.~R.,  {Swain} M.~R.,  {Zellem} R.~T.,  {Line} M.~R.,  {Roudier} G.,
  {Rocha} G.,   {Livingston} J.~H.,  2016, \mn@doi [\apj]
  {10.3847/0004-637X/823/2/109}, \href
  {https://ui.adsabs.harvard.edu/abs/2016ApJ...823..109I} {823, 109}

\bibitem[\protect\citeauthoryear{{Jord{\'a}n} et~al.,}{{Jord{\'a}n}
  et~al.}{2013}]{2013ApJ...778..184J}
{Jord{\'a}n} A.,  et~al., 2013, \mn@doi [\apj] {10.1088/0004-637X/778/2/184},
  \href {https://ui.adsabs.harvard.edu/abs/2013ApJ...778..184J} {778, 184}

\bibitem[\protect\citeauthoryear{{Kawashima} \& {Ikoma}}{{Kawashima} \&
  {Ikoma}}{2019}]{2019ApJ...877..109K}
{Kawashima} Y.,  {Ikoma} M.,  2019, \mn@doi [\apj] {10.3847/1538-4357/ab1b1d},
  \href {https://ui.adsabs.harvard.edu/abs/2019ApJ...877..109K} {877, 109}

\bibitem[\protect\citeauthoryear{{Kesseli}, {West}, {Veyette}, {Harrison},
  {Feldman}  \& {Bochanski}}{{Kesseli} et~al.}{2017}]{2017ApJS..230...16K}
{Kesseli} A.~Y.,  {West} A.~A.,  {Veyette} M.,  {Harrison} B.,  {Feldman} D.,
  {Bochanski} J.~J.,  2017, \mn@doi [\apjs] {10.3847/1538-4365/aa656d}, \href
  {https://ui.adsabs.harvard.edu/abs/2017ApJS..230...16K} {230, 16}

\bibitem[\protect\citeauthoryear{{Kirk}, {Wheatley}, {Louden}, {Doyle},
  {Skillen}, {McCormac}, {Irwin}  \& {Karjalainen}}{{Kirk}
  et~al.}{2017}]{2017MNRAS.468.3907K}
{Kirk} J.,  {Wheatley} P.~J.,  {Louden} T.,  {Doyle} A.~P.,  {Skillen} I.,
  {McCormac} J.,  {Irwin} P.~G.~J.,   {Karjalainen} R.,  2017, \mn@doi [\mnras]
  {10.1093/mnras/stx752}, \href
  {https://ui.adsabs.harvard.edu/abs/2017MNRAS.468.3907K} {468, 3907}

\bibitem[\protect\citeauthoryear{{Kitzmann} \& {Heng}}{{Kitzmann} \&
  {Heng}}{2018}]{2018MNRAS.475...94K}
{Kitzmann} D.,  {Heng} K.,  2018, \mn@doi [\mnras] {10.1093/mnras/stx3141},
  \href {https://ui.adsabs.harvard.edu/abs/2018MNRAS.475...94K} {475, 94}

\bibitem[\protect\citeauthoryear{{Komacek}, {Showman}  \&
  {Parmentier}}{{Komacek} et~al.}{2019}]{2019ApJ...881..152K}
{Komacek} T.~D.,  {Showman} A.~P.,   {Parmentier} V.,  2019, \mn@doi [\apj]
  {10.3847/1538-4357/ab338b}, \href
  {https://ui.adsabs.harvard.edu/abs/2019ApJ...881..152K} {881, 152}

\bibitem[\protect\citeauthoryear{{Kreidberg}}{{Kreidberg}}{2015}]{2015PASP..127.1161K}
{Kreidberg} L.,  2015, \mn@doi [\pasp] {10.1086/683602}, \href
  {https://ui.adsabs.harvard.edu/abs/2015PASP..127.1161K} {127, 1161}

\bibitem[\protect\citeauthoryear{{Kreidberg} et~al.,}{{Kreidberg}
  et~al.}{2014}]{2014ApJ...793L..27K}
{Kreidberg} L.,  et~al., 2014, \mn@doi [\apjl] {10.1088/2041-8205/793/2/L27},
  \href {https://ui.adsabs.harvard.edu/abs/2014ApJ...793L..27K} {793, L27}

\bibitem[\protect\citeauthoryear{{Lecavelier Des Etangs}, {Pont},
  {Vidal-Madjar}  \& {Sing}}{{Lecavelier Des Etangs}
  et~al.}{2008}]{2008A&A...481L..83L}
{Lecavelier Des Etangs} A.,  {Pont} F.,  {Vidal-Madjar} A.,   {Sing} D.,  2008,
  \mn@doi [\aap] {10.1051/0004-6361:200809388}, \href
  {https://ui.adsabs.harvard.edu/abs/2008A&A...481L..83L} {481, L83}

\bibitem[\protect\citeauthoryear{{Lightkurve Collaboration}
  et~al.,}{{Lightkurve Collaboration} et~al.}{2018}]{2018ascl.soft12013L}
{Lightkurve Collaboration} et~al., 2018, {Lightkurve: Kepler and TESS time
  series analysis in Python} (\mn@eprint {ascl} {1812.013})

\bibitem[\protect\citeauthoryear{{Lothringer}, {Fu}, {Sing}  \&
  {Barman}}{{Lothringer} et~al.}{2020}]{2020ApJ...898L..14L}
{Lothringer} J.~D.,  {Fu} G.,  {Sing} D.~K.,   {Barman} T.~S.,  2020, \mn@doi
  [\apjl] {10.3847/2041-8213/aba265}, \href
  {https://ui.adsabs.harvard.edu/abs/2020ApJ...898L..14L} {898, L14}

\bibitem[\protect\citeauthoryear{{Luger}, {Agol}, {Kruse}, {Barnes}, {Becker},
  {Foreman-Mackey}  \& {Deming}}{{Luger} et~al.}{2016}]{2016AJ....152..100L}
{Luger} R.,  {Agol} E.,  {Kruse} E.,  {Barnes} R.,  {Becker} A.,
  {Foreman-Mackey} D.,   {Deming} D.,  2016, \mn@doi [\aj]
  {10.3847/0004-6256/152/4/100}, \href
  {https://ui.adsabs.harvard.edu/abs/2016AJ....152..100L} {152, 100}

\bibitem[\protect\citeauthoryear{{Luger}, {Kruse}, {Foreman-Mackey}, {Agol}  \&
  {Saunders}}{{Luger} et~al.}{2018}]{2018AJ....156...99L}
{Luger} R.,  {Kruse} E.,  {Foreman-Mackey} D.,  {Agol} E.,   {Saunders} N.,
  2018, \mn@doi [\aj] {10.3847/1538-3881/aad230}, \href
  {https://ui.adsabs.harvard.edu/abs/2018AJ....156...99L} {156, 99}

\bibitem[\protect\citeauthoryear{{Madhusudhan}}{{Madhusudhan}}{2019}]{2019ARA&A..57..617M}
{Madhusudhan} N.,  2019, \mn@doi [\araa] {10.1146/annurev-astro-081817-051846},
  \href {https://ui.adsabs.harvard.edu/abs/2019ARA&A..57..617M} {57, 617}

\bibitem[\protect\citeauthoryear{{Mallonn} \& {Strassmeier}}{{Mallonn} \&
  {Strassmeier}}{2016}]{2016A&A...590A.100M}
{Mallonn} M.,  {Strassmeier} K.~G.,  2016, \mn@doi [\aap]
  {10.1051/0004-6361/201527898}, \href
  {https://ui.adsabs.harvard.edu/abs/2016A&A...590A.100M} {590, A100}

\bibitem[\protect\citeauthoryear{{Mandel} \& {Agol}}{{Mandel} \&
  {Agol}}{2002}]{2002ApJ...580L.171M}
{Mandel} K.,  {Agol} E.,  2002, \mn@doi [\apjl] {10.1086/345520}, \href
  {https://ui.adsabs.harvard.edu/abs/2002ApJ...580L.171M} {580, L171}

\bibitem[\protect\citeauthoryear{{May}, {Gardner}, {Rauscher}  \&
  {Monnier}}{{May} et~al.}{2020}]{2020AJ....159....7M}
{May} E.~M.,  {Gardner} T.,  {Rauscher} E.,   {Monnier} J.~D.,  2020, \mn@doi
  [\aj] {10.3847/1538-3881/ab5361}, \href
  {https://ui.adsabs.harvard.edu/abs/2020AJ....159....7M} {159, 7}

\bibitem[\protect\citeauthoryear{{McCullough}, {Crouzet}, {Deming}  \&
  {Madhusudhan}}{{McCullough} et~al.}{2014}]{2014ApJ...791...55M}
{McCullough} P.~R.,  {Crouzet} N.,  {Deming} D.,   {Madhusudhan} N.,  2014,
  \mn@doi [\apj] {10.1088/0004-637X/791/1/55}, \href
  {https://ui.adsabs.harvard.edu/abs/2014ApJ...791...55M} {791, 55}

\bibitem[\protect\citeauthoryear{{Moffat}}{{Moffat}}{1969}]{1969A&A.....3..455M}
{Moffat} A.~F.~J.,  1969, \aap, \href
  {https://ui.adsabs.harvard.edu/abs/1969A&A.....3..455M} {3, 455}

\bibitem[\protect\citeauthoryear{{Molli{\`e}re}, {Wardenier}, {van Boekel},
  {Henning}, {Molaverdikhani}  \& {Snellen}}{{Molli{\`e}re}
  et~al.}{2019}]{2019A&A...627A..67M}
{Molli{\`e}re} P.,  {Wardenier} J.~P.,  {van Boekel} R.,  {Henning} T.,
  {Molaverdikhani} K.,   {Snellen} I.~A.~G.,  2019, \mn@doi [\aap]
  {10.1051/0004-6361/201935470}, \href
  {https://ui.adsabs.harvard.edu/abs/2019A&A...627A..67M} {627, A67}

\bibitem[\protect\citeauthoryear{{Morris}}{{Morris}}{2020}]{2020ApJ...893...67M}
{Morris} B.~M.,  2020, \mn@doi [\apj] {10.3847/1538-4357/ab79a0}, \href
  {https://ui.adsabs.harvard.edu/abs/2020ApJ...893...67M} {893, 67}

\bibitem[\protect\citeauthoryear{{Mo{\v{c}}nik}, {Hellier}  \&
  {Southworth}}{{Mo{\v{c}}nik} et~al.}{2018}]{2018AJ....156...44M}
{Mo{\v{c}}nik} T.,  {Hellier} C.,   {Southworth} J.,  2018, \mn@doi [\aj]
  {10.3847/1538-3881/aacb26}, \href
  {https://ui.adsabs.harvard.edu/abs/2018AJ....156...44M} {156, 44}

\bibitem[\protect\citeauthoryear{{Murgas}, {Chen}, {Pall{\'e}}, {Nortmann}  \&
  {Nowak}}{{Murgas} et~al.}{2019}]{2019A&A...622A.172M}
{Murgas} F.,  {Chen} G.,  {Pall{\'e}} E.,  {Nortmann} L.,   {Nowak} G.,  2019,
  \mn@doi [\aap] {10.1051/0004-6361/201834063}, \href
  {https://ui.adsabs.harvard.edu/abs/2019A&A...622A.172M} {622, A172}

\bibitem[\protect\citeauthoryear{{Murgas}, {Chen}, {Nortmann}, {Palle}  \&
  {Nowak}}{{Murgas} et~al.}{2020}]{2020A&A...641A.158M}
{Murgas} F.,  {Chen} G.,  {Nortmann} L.,  {Palle} E.,   {Nowak} G.,  2020,
  \mn@doi [\aap] {10.1051/0004-6361/202038161}, \href
  {https://ui.adsabs.harvard.edu/abs/2020A&A...641A.158M} {641, A158}

\bibitem[\protect\citeauthoryear{{Narita} et~al.,}{{Narita}
  et~al.}{2019}]{2019JATIS...5a5001N}
{Narita} N.,  et~al., 2019, \mn@doi [Journal of Astronomical Telescopes,
  Instruments, and Systems] {10.1117/1.JATIS.5.1.015001}, \href
  {https://ui.adsabs.harvard.edu/abs/2019JATIS...5a5001N} {5, 015001}

\bibitem[\protect\citeauthoryear{{Nikolov} et~al.,}{{Nikolov}
  et~al.}{2015}]{2015MNRAS.447..463N}
{Nikolov} N.,  et~al., 2015, \mn@doi [\mnras] {10.1093/mnras/stu2433}, \href
  {https://ui.adsabs.harvard.edu/abs/2015MNRAS.447..463N} {447, 463}

\bibitem[\protect\citeauthoryear{{Nikolov} et~al.,}{{Nikolov}
  et~al.}{2018}]{2018Natur.557..526N}
{Nikolov} N.,  et~al., 2018, \mn@doi [\nat] {10.1038/s41586-018-0101-7}, \href
  {https://ui.adsabs.harvard.edu/abs/2018Natur.557..526N} {557, 526}

\bibitem[\protect\citeauthoryear{{Ochsenbein}, {Bauer}  \&
  {Marcout}}{{Ochsenbein} et~al.}{2000}]{2000A&AS..143...23O}
{Ochsenbein} F.,  {Bauer} P.,   {Marcout} J.,  2000, \mn@doi [\aaps]
  {10.1051/aas:2000169}, \href
  {https://ui.adsabs.harvard.edu/abs/2000A&AS..143...23O} {143, 23}

\bibitem[\protect\citeauthoryear{{Ohno} \& {Kawashima}}{{Ohno} \&
  {Kawashima}}{2020}]{2020ApJ...895L..47O}
{Ohno} K.,  {Kawashima} Y.,  2020, \mn@doi [\apjl] {10.3847/2041-8213/ab93d7},
  \href {https://ui.adsabs.harvard.edu/abs/2020ApJ...895L..47O} {895, L47}

\bibitem[\protect\citeauthoryear{{Oke} \& {Gunn}}{{Oke} \&
  {Gunn}}{1982}]{1982PASP...94..586O}
{Oke} J.~B.,  {Gunn} J.~E.,  1982, \mn@doi [\pasp] {10.1086/131027}, \href
  {https://ui.adsabs.harvard.edu/abs/1982PASP...94..586O} {94, 586}

\bibitem[\protect\citeauthoryear{{Oshagh}, {Santos}, {Ehrenreich},
  {Haghighipour}, {Figueira}, {Santerne}  \& {Montalto}}{{Oshagh}
  et~al.}{2014}]{2014A&A...568A..99O}
{Oshagh} M.,  {Santos} N.~C.,  {Ehrenreich} D.,  {Haghighipour} N.,  {Figueira}
  P.,  {Santerne} A.,   {Montalto} M.,  2014, \mn@doi [\aap]
  {10.1051/0004-6361/201424059}, \href
  {https://ui.adsabs.harvard.edu/abs/2014A&A...568A..99O} {568, A99}

\bibitem[\protect\citeauthoryear{{Parviainen}, {Pall{\'e}}, {Nortmann},
  {Nowak}, {Iro}, {Murgas}  \& {Aigrain}}{{Parviainen}
  et~al.}{2016}]{2016A&A...585A.114P}
{Parviainen} H.,  {Pall{\'e}} E.,  {Nortmann} L.,  {Nowak} G.,  {Iro} N.,
  {Murgas} F.,   {Aigrain} S.,  2016, \mn@doi [\aap]
  {10.1051/0004-6361/201526313}, \href
  {https://ui.adsabs.harvard.edu/abs/2016A&A...585A.114P} {585, A114}

\bibitem[\protect\citeauthoryear{{Pearson}, {Griffith}, {Zellem}, {Koskinen}
  \& {Roudier}}{{Pearson} et~al.}{2019}]{2019AJ....157...21P}
{Pearson} K.~A.,  {Griffith} C.~A.,  {Zellem} R.~T.,  {Koskinen} T.~T.,
  {Roudier} G.~M.,  2019, \mn@doi [\aj] {10.3847/1538-3881/aaf1ae}, \href
  {https://ui.adsabs.harvard.edu/abs/2019AJ....157...21P} {157, 21}

\bibitem[\protect\citeauthoryear{{Pinhas} \& {Madhusudhan}}{{Pinhas} \&
  {Madhusudhan}}{2017}]{2017MNRAS.471.4355P}
{Pinhas} A.,  {Madhusudhan} N.,  2017, \mn@doi [\mnras]
  {10.1093/mnras/stx1849}, \href
  {https://ui.adsabs.harvard.edu/abs/2017MNRAS.471.4355P} {471, 4355}

\bibitem[\protect\citeauthoryear{{Pinhas}, {Madhusudhan}, {Gandhi}  \&
  {MacDonald}}{{Pinhas} et~al.}{2019}]{2019MNRAS.482.1485P}
{Pinhas} A.,  {Madhusudhan} N.,  {Gandhi} S.,   {MacDonald} R.,  2019, \mn@doi
  [\mnras] {10.1093/mnras/sty2544}, \href
  {https://ui.adsabs.harvard.edu/abs/2019MNRAS.482.1485P} {482, 1485}

\bibitem[\protect\citeauthoryear{{Pont}, {Sing}, {Gibson}, {Aigrain}, {Henry}
  \& {Husnoo}}{{Pont} et~al.}{2013}]{2013MNRAS.432.2917P}
{Pont} F.,  {Sing} D.~K.,  {Gibson} N.~P.,  {Aigrain} S.,  {Henry} G.,
  {Husnoo} N.,  2013, \mn@doi [\mnras] {10.1093/mnras/stt651}, \href
  {https://ui.adsabs.harvard.edu/abs/2013MNRAS.432.2917P} {432, 2917}

\bibitem[\protect\citeauthoryear{{Rackham}, {Apai}  \& {Giampapa}}{{Rackham}
  et~al.}{2018}]{2018ApJ...853..122R}
{Rackham} B.~V.,  {Apai} D.,   {Giampapa} M.~S.,  2018, \mn@doi [\apj]
  {10.3847/1538-4357/aaa08c}, \href
  {https://ui.adsabs.harvard.edu/abs/2018ApJ...853..122R} {853, 122}

\bibitem[\protect\citeauthoryear{{Rackham}, {Apai}  \& {Giampapa}}{{Rackham}
  et~al.}{2019}]{2019AJ....157...96R}
{Rackham} B.~V.,  {Apai} D.,   {Giampapa} M.~S.,  2019, \mn@doi [\aj]
  {10.3847/1538-3881/aaf892}, \href
  {https://ui.adsabs.harvard.edu/abs/2019AJ....157...96R} {157, 96}

\bibitem[\protect\citeauthoryear{{Rasmussen} \& {Williams}}{{Rasmussen} \&
  {Williams}}{2006}]{2006gpml.book.....R}
{Rasmussen} C.~E.,  {Williams} C. K.~I.,  2006, {Gaussian Processes for Machine
  Learning}

\bibitem[\protect\citeauthoryear{{Rosich}, {Herrero}, {Mallonn}, {Ribas},
  {Morales}, {Perger}, {Anglada-Escud{\'e}}  \& {Granzer}}{{Rosich}
  et~al.}{2020}]{2020A&A...641A..82R}
{Rosich} A.,  {Herrero} E.,  {Mallonn} M.,  {Ribas} I.,  {Morales} J.~C.,
  {Perger} M.,  {Anglada-Escud{\'e}} G.,   {Granzer} T.,  2020, \mn@doi [\aap]
  {10.1051/0004-6361/202037586}, \href
  {https://ui.adsabs.harvard.edu/abs/2020A&A...641A..82R} {641, A82}

\bibitem[\protect\citeauthoryear{{Schlafly} \& {Finkbeiner}}{{Schlafly} \&
  {Finkbeiner}}{2011}]{2011ApJ...737..103S}
{Schlafly} E.~F.,  {Finkbeiner} D.~P.,  2011, \mn@doi [\apj]
  {10.1088/0004-637X/737/2/103}, \href
  {https://ui.adsabs.harvard.edu/abs/2011ApJ...737..103S} {737, 103}

\bibitem[\protect\citeauthoryear{Schwarz}{Schwarz}{1978}]{schwarz1978}
Schwarz G.,  1978, \mn@doi [Ann. Statist.] {10.1214/aos/1176344136}, 6, 461

\bibitem[\protect\citeauthoryear{{Seager} \& {Mall{\'e}n-Ornelas}}{{Seager} \&
  {Mall{\'e}n-Ornelas}}{2003}]{2003ApJ...585.1038S}
{Seager} S.,  {Mall{\'e}n-Ornelas} G.,  2003, \mn@doi [\apj] {10.1086/346105},
  \href {https://ui.adsabs.harvard.edu/abs/2003ApJ...585.1038S} {585, 1038}

\bibitem[\protect\citeauthoryear{{Seager} \& {Sasselov}}{{Seager} \&
  {Sasselov}}{2000}]{2000ApJ...537..916S}
{Seager} S.,  {Sasselov} D.~D.,  2000, \mn@doi [\apj] {10.1086/309088}, \href
  {https://ui.adsabs.harvard.edu/abs/2000ApJ...537..916S} {537, 916}

\bibitem[\protect\citeauthoryear{{Sedaghati} et~al.,}{{Sedaghati}
  et~al.}{2017}]{2017Natur.549..238S}
{Sedaghati} E.,  et~al., 2017, \mn@doi [\nat] {10.1038/nature23651}, \href
  {https://ui.adsabs.harvard.edu/abs/2017Natur.549..238S} {549, 238}

\bibitem[\protect\citeauthoryear{{Sing} et~al.,}{{Sing}
  et~al.}{2011}]{2011MNRAS.416.1443S}
{Sing} D.~K.,  et~al., 2011, \mn@doi [\mnras]
  {10.1111/j.1365-2966.2011.19142.x}, \href
  {https://ui.adsabs.harvard.edu/abs/2011MNRAS.416.1443S} {416, 1443}

\bibitem[\protect\citeauthoryear{{Sing} et~al.,}{{Sing}
  et~al.}{2013}]{2013MNRAS.436.2956S}
{Sing} D.~K.,  et~al., 2013, \mn@doi [\mnras] {10.1093/mnras/stt1782}, \href
  {https://ui.adsabs.harvard.edu/abs/2013MNRAS.436.2956S} {436, 2956}

\bibitem[\protect\citeauthoryear{{Sing} et~al.,}{{Sing}
  et~al.}{2016}]{2016Natur.529...59S}
{Sing} D.~K.,  et~al., 2016, \mn@doi [\nat] {10.1038/nature16068}, \href
  {https://ui.adsabs.harvard.edu/abs/2016Natur.529...59S} {529, 59}

\bibitem[\protect\citeauthoryear{{Smith} et~al.,}{{Smith}
  et~al.}{2014}]{2014A&A...570A..64S}
{Smith} A.~M.~S.,  et~al., 2014, \mn@doi [\aap] {10.1051/0004-6361/201424752},
  \href {https://ui.adsabs.harvard.edu/abs/2014A&A...570A..64S} {570, A64}

\bibitem[\protect\citeauthoryear{{Sousa} et~al.,}{{Sousa}
  et~al.}{2018}]{2018A&A...620A..58S}
{Sousa} S.~G.,  et~al., 2018, \mn@doi [\aap] {10.1051/0004-6361/201833350},
  \href {https://ui.adsabs.harvard.edu/abs/2018A&A...620A..58S} {620, A58}

\bibitem[\protect\citeauthoryear{{Southworth}}{{Southworth}}{2011}]{2011MNRAS.417.2166S}
{Southworth} J.,  2011, \mn@doi [\mnras] {10.1111/j.1365-2966.2011.19399.x},
  \href {https://ui.adsabs.harvard.edu/abs/2011MNRAS.417.2166S} {417, 2166}

\bibitem[\protect\citeauthoryear{{Stevenson}}{{Stevenson}}{2016}]{2016ApJ...817L..16S}
{Stevenson} K.~B.,  2016, \mn@doi [\apjl] {10.3847/2041-8205/817/2/L16}, \href
  {https://ui.adsabs.harvard.edu/abs/2016ApJ...817L..16S} {817, L16}

\bibitem[\protect\citeauthoryear{{Tody}}{{Tody}}{1993}]{1993ASPC...52..173T}
{Tody} D.,  1993, in {Hanisch} R.~J.,  {Brissenden} R.~J.~V.,   {Barnes} J.,
  eds,  Astronomical Society of the Pacific Conference Series Vol. 52,
  Astronomical Data Analysis Software and Systems II. p.~173

\bibitem[\protect\citeauthoryear{{Tsiaras} et~al.,}{{Tsiaras}
  et~al.}{2018}]{2018AJ....155..156T}
{Tsiaras} A.,  et~al., 2018, \mn@doi [\aj] {10.3847/1538-3881/aaaf75}, \href
  {https://ui.adsabs.harvard.edu/abs/2018AJ....155..156T} {155, 156}

\bibitem[\protect\citeauthoryear{{Wakeford} \& {Sing}}{{Wakeford} \&
  {Sing}}{2015}]{2015A&A...573A.122W}
{Wakeford} H.~R.,  {Sing} D.~K.,  2015, \mn@doi [\aap]
  {10.1051/0004-6361/201424207}, \href
  {https://ui.adsabs.harvard.edu/abs/2015A&A...573A.122W} {573, A122}

\bibitem[\protect\citeauthoryear{{Wakeford} et~al.,}{{Wakeford}
  et~al.}{2017a}]{2017Sci...356..628W}
{Wakeford} H.~R.,  et~al., 2017a, \mn@doi [Science] {10.1126/science.aah4668},
  \href {https://ui.adsabs.harvard.edu/abs/2017Sci...356..628W} {356, 628}

\bibitem[\protect\citeauthoryear{{Wakeford}, {Visscher}, {Lewis}, {Kataria},
  {Marley}, {Fortney}  \& {Mand ell}}{{Wakeford}
  et~al.}{2017b}]{2017MNRAS.464.4247W}
{Wakeford} H.~R.,  {Visscher} C.,  {Lewis} N.~K.,  {Kataria} T.,  {Marley}
  M.~S.,  {Fortney} J.~J.,   {Mand ell} A.~M.,  2017b, \mn@doi [\mnras]
  {10.1093/mnras/stw2639}, \href
  {https://ui.adsabs.harvard.edu/abs/2017MNRAS.464.4247W} {464, 4247}

\bibitem[\protect\citeauthoryear{{Weaver} et~al.,}{{Weaver}
  et~al.}{2020}]{2020AJ....159...13W}
{Weaver} I.~C.,  et~al., 2020, \mn@doi [\aj] {10.3847/1538-3881/ab55da}, \href
  {https://ui.adsabs.harvard.edu/abs/2020AJ....159...13W} {159, 13}

\bibitem[\protect\citeauthoryear{{Welbanks}, {Madhusudhan}, {Allard}, {Hubeny},
  {Spiegelman}  \& {Leininger}}{{Welbanks} et~al.}{2019}]{2019ApJ...887L..20W}
{Welbanks} L.,  {Madhusudhan} N.,  {Allard} N.~F.,  {Hubeny} I.,  {Spiegelman}
  F.,   {Leininger} T.,  2019, \mn@doi [\apjl] {10.3847/2041-8213/ab5a89},
  \href {https://ui.adsabs.harvard.edu/abs/2019ApJ...887L..20W} {887, L20}

\bibitem[\protect\citeauthoryear{{Woitke}, {Helling}, {Hunter}, {Millard},
  {Turner}, {Worters}, {Blecic}  \& {Stock}}{{Woitke}
  et~al.}{2018}]{2018A&A...614A...1W}
{Woitke} P.,  {Helling} C.,  {Hunter} G.~H.,  {Millard} J.~D.,  {Turner} G.~E.,
   {Worters} M.,  {Blecic} J.,   {Stock} J.~W.,  2018, \mn@doi [\aap]
  {10.1051/0004-6361/201732193}, \href
  {https://ui.adsabs.harvard.edu/abs/2018A&A...614A...1W} {614, A1}

\bibitem[\protect\citeauthoryear{{Wong} et~al.,}{{Wong}
  et~al.}{2020}]{2020AJ....159..234W}
{Wong} I.,  et~al., 2020, \mn@doi [\aj] {10.3847/1538-3881/ab880d}, \href
  {https://ui.adsabs.harvard.edu/abs/2020AJ....159..234W} {159, 234}

\bibitem[\protect\citeauthoryear{{Zacharias} et~al.,}{{Zacharias}
  et~al.}{2015}]{2015AJ....150..101Z}
{Zacharias} N.,  et~al., 2015, \mn@doi [\aj] {10.1088/0004-6256/150/4/101},
  \href {https://ui.adsabs.harvard.edu/abs/2015AJ....150..101Z} {150, 101}

\bibitem[\protect\citeauthoryear{{Zahnle}, {Marley}, {Freedman}, {Lodders}  \&
  {Fortney}}{{Zahnle} et~al.}{2009}]{2009ApJ...701L..20Z}
{Zahnle} K.,  {Marley} M.~S.,  {Freedman} R.~S.,  {Lodders} K.,   {Fortney}
  J.~J.,  2009, \mn@doi [\apjl] {10.1088/0004-637X/701/1/L20}, \href
  {https://ui.adsabs.harvard.edu/abs/2009ApJ...701L..20Z} {701, L20}

\bibitem[\protect\citeauthoryear{{Zhang}}{{Zhang}}{2020}]{2020RAA....20..099Z}
{Zhang} X.,  2020, \mn@doi [Research in Astronomy and Astrophysics]
  {10.1088/1674-4527/20/7/99}, \href
  {https://ui.adsabs.harvard.edu/abs/2020RAA....20..099Z} {20, 099}

\bibitem[\protect\citeauthoryear{{Zhang}, {Chachan}, {Kempton}  \&
  {Knutson}}{{Zhang} et~al.}{2019}]{2019PASP..131c4501Z}
{Zhang} M.,  {Chachan} Y.,  {Kempton} E. M.~R.,   {Knutson} H.~A.,  2019,
  \mn@doi [\pasp] {10.1088/1538-3873/aaf5ad}, \href
  {https://ui.adsabs.harvard.edu/abs/2019PASP..131c4501Z} {131, 034501}

\bibitem[\protect\citeauthoryear{{Zhang}, {Chachan}, {Kempton}, {Knutson}  \&
  {Chang}}{{Zhang} et~al.}{2020}]{2020ApJ...899...27Z}
{Zhang} M.,  {Chachan} Y.,  {Kempton} E. M.~R.,  {Knutson} H.~A.,   {Chang}
  W.~H.,  2020, \mn@doi [\apj] {10.3847/1538-4357/aba1e6}, \href
  {https://ui.adsabs.harvard.edu/abs/2020ApJ...899...27Z} {899, 27}

\makeatother
\end{thebibliography}

\section*{Affiliations}
{\it
~$^{1}$Key Laboratory of Planetary Sciences, Purple Mountain Observatory, Chinese Academy of Sciences, Nanjing 210023, PR China\\ 
~$^{2}$Instituto de Astrof\'{i}sica de Canarias, V\'{i}a L\'{a}ctea s/n, E-38205 La Laguna, Tenerife, Spain\\ 
~$^{3}$Departamento de Astrof\'{i}sica, Universidad de La Laguna, Spain\\ 
~$^{4}$Key Laboratory of Radio Astronomy, Purple Mountain Observatory, Chinese Academy of Sciences, Nanjing 210023, PR China\\ 
~$^{5}$Max-Planck-Institut f\"{u}r Astronomie, K\"{o}nigstuhl 17, 69117 Heidelberg, Germany\\ 
~$^{6}$Institut f\"{u}r Astrophysik, Georg-August-Universit\"{a}t, Friedrich-Hund-Platz 1, D-37077 G\"{o}ttingen, Germany\\ 
~$^{7}$Science Support Office, Directorate of Science, European Space Research and Technology Centre (ESA/ESTEC), Keplerlaan 1, 2201 AZ Noordwijk, The Netherlands\\ 
~$^{8}$Department of Earth and Planetary Science, Graduate School of Science, The University of Tokyo, Japan\\ 
~$^{9}$MTA-ELTE Exoplanet Research Group, 9700 Szombathely, Szent Imre h.~u.~112, Hungary \\ 
$^{10}$ELTE Gothard Astrophysical Observatory, 9700 Szombathely, Szent Imre h.~u.~112, Hungary \\ 
$^{11}$Astronomical Institute, Slovak Academy of Sciences, 05960 Tatransk\'{a} Lomnica, Slovakia \\
$^{12}$Institute of Planetary Research, German Aerospace Center, Rutherfordstrasse 2, 12489 Berlin, Germany\\ 
$^{13}$Astrobiology Center of NINS, 2-21-1, Osawa, Mitaka, Tokyo 181-8588, Japan\\ 
$^{14}$National Astronomical Observatory of Japan, 2-21-1 Osawa, Mitaka, Tokyo 181-8588, Japan\\ 
$^{15}$Department of Astronomy, The University of Tokyo, 7-3-1 Hongo, Bunkyo-ku, Tokyo 113-0033, Japan\\ 
$^{16}$Komaba Institute for Science, The University of Tokyo, 3-8-1 Komaba, Meguro, Tokyo 153-8902, Japan\\ 
$^{17}$Japan Science and Technology Agency, PRESTO, 3-8-1 Komaba, Meguro, Tokyo 153-8902, Japan\\ 
$^{18}$Department of Astronomical Science, The Graduated University for Advanced Studies, SOKENDAI, 2-21-1, Osawa, Mitaka, Tokyo, 181-8588 Japan\\ 
$^{19}$Institute of Astronomy and Astrophysics, Academia Sinica, P.O. Box 23-141, Taipei 10617, Taiwan, R.O.C.\\
$^{20}$Department of Astrophysics, National Taiwan University, Taipei 10617, Taiwan, R.O.C.
}




\appendix

\section{Additional tables and figures}
%

\begin{table}
\scriptsize
\centering
\caption{Mid-transit times of WASP-104b.}
\label{tab:tmid}
\begin{tabular}{cccc}
\hline\hline
Epoch & Mid-transit time & Residuals & Instrument\\
& ($\mathrm{BJD}_\mathrm{TDB}-2450000$) & (sec) & \\
\hline
-885 & 6381.536026 $\pm$ 0.000406 & -19.0 & TRAPPIST\\
-881 & 6388.558002 $\pm$ 0.000786 & 11.5 & TRAPPIST\\
-877 & 6395.578120 $\pm$ 0.000751 & -118.5 & TRAPPIST\\
-873 & 6402.601347 $\pm$ 0.000192 & 20.1 & Euler\\
-869 & 6409.622108 $\pm$ 0.000457 & -54.3 & Euler\\
-848 & 6446.485884 $\pm$ 0.000328 & -32.1 & TRAPPIST\\
-494 & 7067.899763 $\pm$ 0.000149 & -7.3 & DBSP\\
-16 & 7906.983803 $\pm$ 0.000069 & 5.6 & {\it K2}\\
-15 & 7908.739168 $\pm$ 0.000069 & 2.1 & {\it K2}\\
-14 & 7910.494656 $\pm$ 0.000058 & 9.2 & {\it K2}\\
-13 & 7912.250033 $\pm$ 0.000058 & 6.7 & {\it K2}\\
-12 & 7914.005334 $\pm$ 0.000056 & -2.3 & {\it K2}\\
-11 & 7915.760811 $\pm$ 0.000061 & 3.9 & {\it K2}\\
-10 & 7917.516096 $\pm$ 0.000063 & -6.6 & {\it K2}\\
 -9 & 7919.271501 $\pm$ 0.000058 & -6.6 & {\it K2}\\
 -8 & 7921.026996 $\pm$ 0.000063 & 1.1 & {\it K2}\\
 -7 & 7922.782210 $\pm$ 0.000070 & -15.4 & {\it K2}\\
 -6 & 7924.537710 $\pm$ 0.000115 & -7.3 & {\it K2}\\
 -5 & 7926.293033 $\pm$ 0.000072 & -14.4 & {\it K2}\\
 -4 & 7928.048514 $\pm$ 0.000063 & -8.0 & {\it K2}\\
 -3 & 7929.803990 $\pm$ 0.000060 & -1.8 & {\it K2}\\
 -2 & 7931.559450 $\pm$ 0.000074 & 2.8 & {\it K2}\\
 -1 & 7933.314857 $\pm$ 0.000064 & 3.0 & {\it K2}\\
  0 & 7935.070336 $\pm$ 0.000100 & 9.3 & {\it K2}\\
  1 & 7936.825603 $\pm$ 0.000057 & -2.7 & {\it K2}\\
  2 & 7938.580941 $\pm$ 0.000087 & -8.5 & {\it K2}\\
  3 & 7940.336465 $\pm$ 0.000073 & 1.7 & {\it K2}\\
  4 & 7942.091794 $\pm$ 0.000062 & -4.9 & {\it K2}\\
  5 & 7943.847253 $\pm$ 0.000081 & -0.3 & {\it K2}\\
  6 & 7945.602633 $\pm$ 0.000057 & -2.5 & {\it K2}\\
  7 & 7947.358142 $\pm$ 0.000072 & 6.4 & {\it K2}\\
  8 & 7949.113446 $\pm$ 0.000072 & -2.4 & {\it K2}\\
  9 & 7950.868920 $\pm$ 0.000057 & 3.6 & {\it K2}\\
 10 & 7952.624273 $\pm$ 0.000065 & -1.0 & {\it K2}\\
 11 & 7954.379661 $\pm$ 0.000057 & -2.5 & {\it K2}\\
 12 & 7956.135168 $\pm$ 0.000060 & 6.2 & {\it K2}\\
 13 & 7957.890516 $\pm$ 0.000060 & 1.2 & {\it K2}\\
 14 & 7959.645980 $\pm$ 0.000070 & 6.4 & {\it K2}\\
 15 & 7961.401285 $\pm$ 0.000082 & -2.4 & {\it K2}\\
 16 & 7963.156847 $\pm$ 0.000060 & 11.1 & {\it K2}\\
 17 & 7964.912183 $\pm$ 0.000071 & 5.1 & {\it K2}\\
 18 & 7966.667603 $\pm$ 0.000071 & 6.3 & {\it K2}\\
 19 & 7968.422901 $\pm$ 0.000070 & -3.0 & {\it K2}\\
 20 & 7970.178242 $\pm$ 0.000071 & -8.5 & {\it K2}\\
 21 & 7971.933684 $\pm$ 0.000084 & -5.4 & {\it K2}\\
 22 & 7973.689231 $\pm$ 0.000071 & 6.8 & {\it K2}\\
 23 & 7975.444730 $\pm$ 0.000065 & 14.9 & {\it K2}\\
 24 & 7977.199945 $\pm$ 0.000090 & -1.5 & {\it K2}\\
 25 & 7978.955214 $\pm$ 0.000080 & -13.4 & {\it K2}\\
 26 & 7980.710816 $\pm$ 0.000081 & 3.5 & {\it K2}\\
 27 & 7982.466215 $\pm$ 0.000063 & 3.0 & {\it K2}\\
 28 & 7984.221461 $\pm$ 0.000062 & -10.8 & {\it K2}\\
112 & 8131.675595 $\pm$ 0.000131 & -5.5 & MuSCAT2\\
120 & 8145.719154 $\pm$ 0.000151 & 21.6 & MuSCAT2\\
162 & 8219.445487 $\pm$ 0.000537 & -39.2 & MuSCAT2\\
166 & 8226.467854 $\pm$ 0.000120 & 25.2 & MuSCAT2\\
296 & 8454.669836 $\pm$ 0.000204 & -39.6 & MuSCAT2\\
300 & 8461.691865 $\pm$ 0.000191 & -4.5 & MuSCAT2\\
304 & 8468.713529 $\pm$ 0.000114 & -0.9 & MuSCAT2\\
308 & 8475.735175 $\pm$ 0.000115 & 1.1 & MuSCAT2\\
329 & 8512.598795 $\pm$ 0.000118 & 10.0 & MuSCAT2\\
391 & 8621.433868 $\pm$ 0.000225 & 3.4 & MuSCAT2\\
546 & 8893.521531 $\pm$ 0.001371 & -14.7 & MuSCAT2\\
558 & 8914.586463 $\pm$ 0.000148 & -9.2 & MuSCAT2\\
562 & 8921.608071 $\pm$ 0.000147 & -10.4 & MuSCAT2\\
567 & 8930.384955 $\pm$ 0.000211 & -22.9 & MuSCAT2\\
\hline
\end{tabular}
\end{table}

\begin{figure}
\centering
\includegraphics[width=\columnwidth]{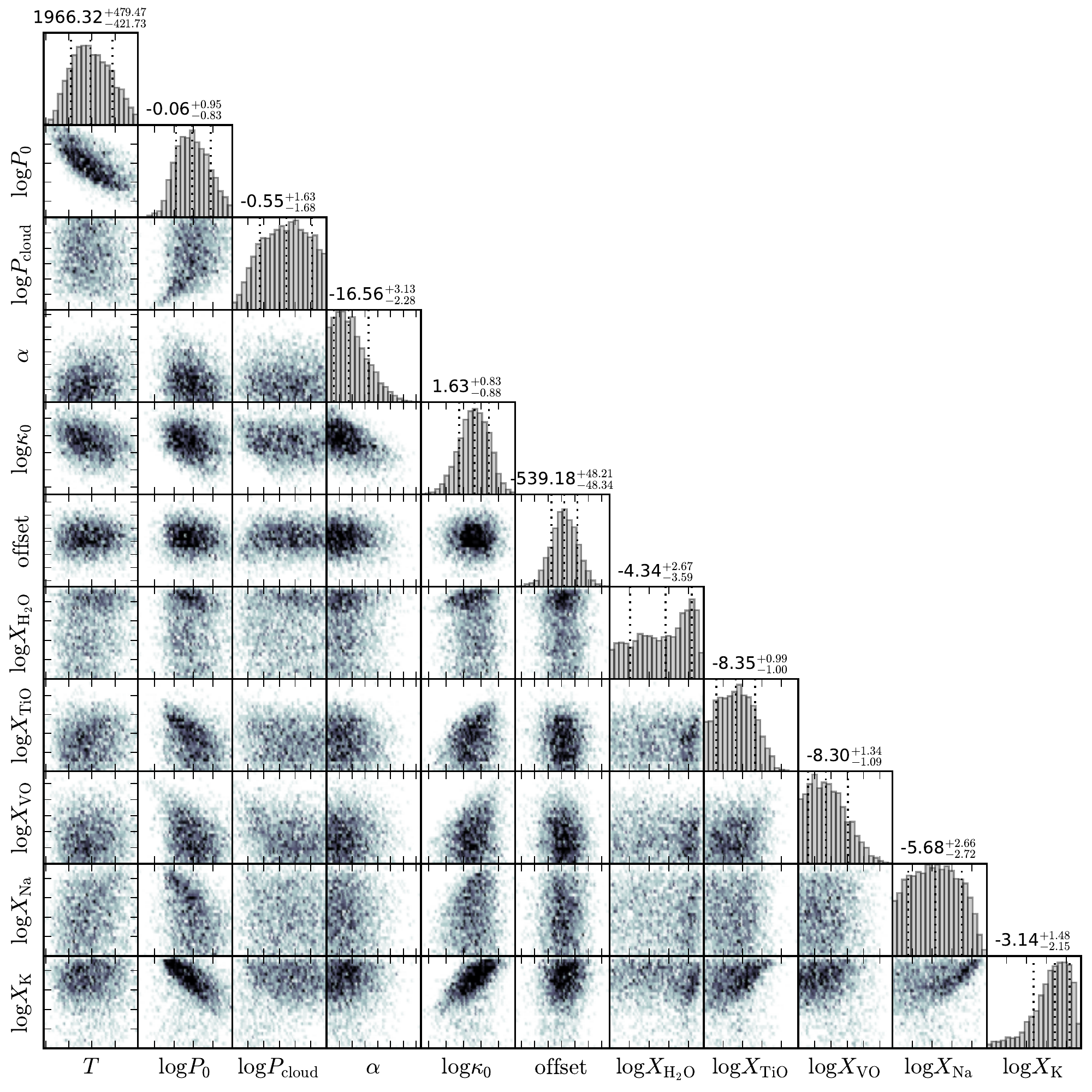}
\caption{Corner plot for the spectral retrieval analysis performed with \texttt{petitRADTRANS}, where no spot contamination is considered.}
\label{fig:corner_petit}
\end{figure}

\begin{figure}
\centering
\includegraphics[width=\columnwidth]{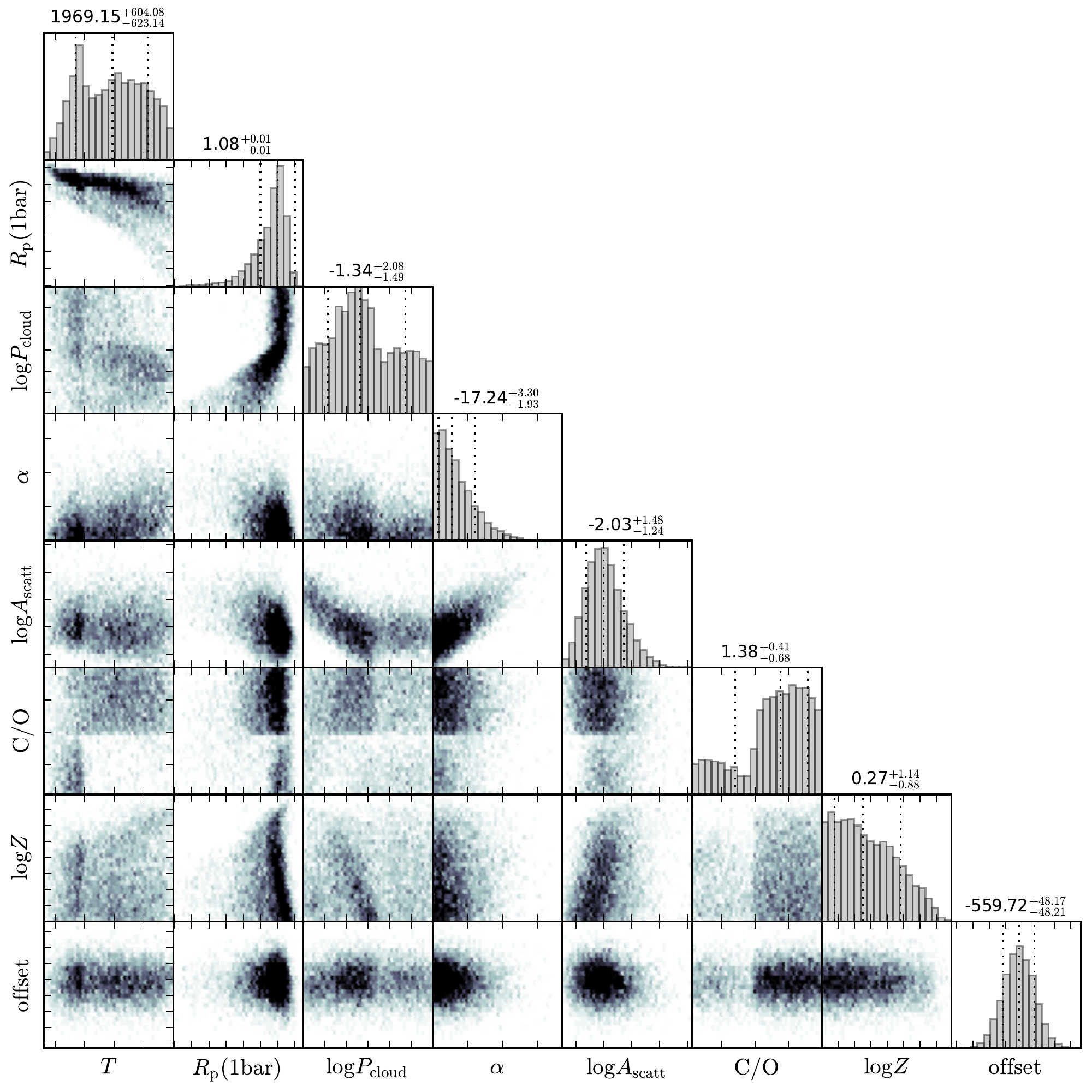}
\caption{Corner plot for the spectral retrieval analysis performed with \texttt{PLATON}, where no spot contamination is considered.}
\label{fig:corner_platon_nospot}
\end{figure}

\begin{figure}
\centering
\includegraphics[width=\columnwidth]{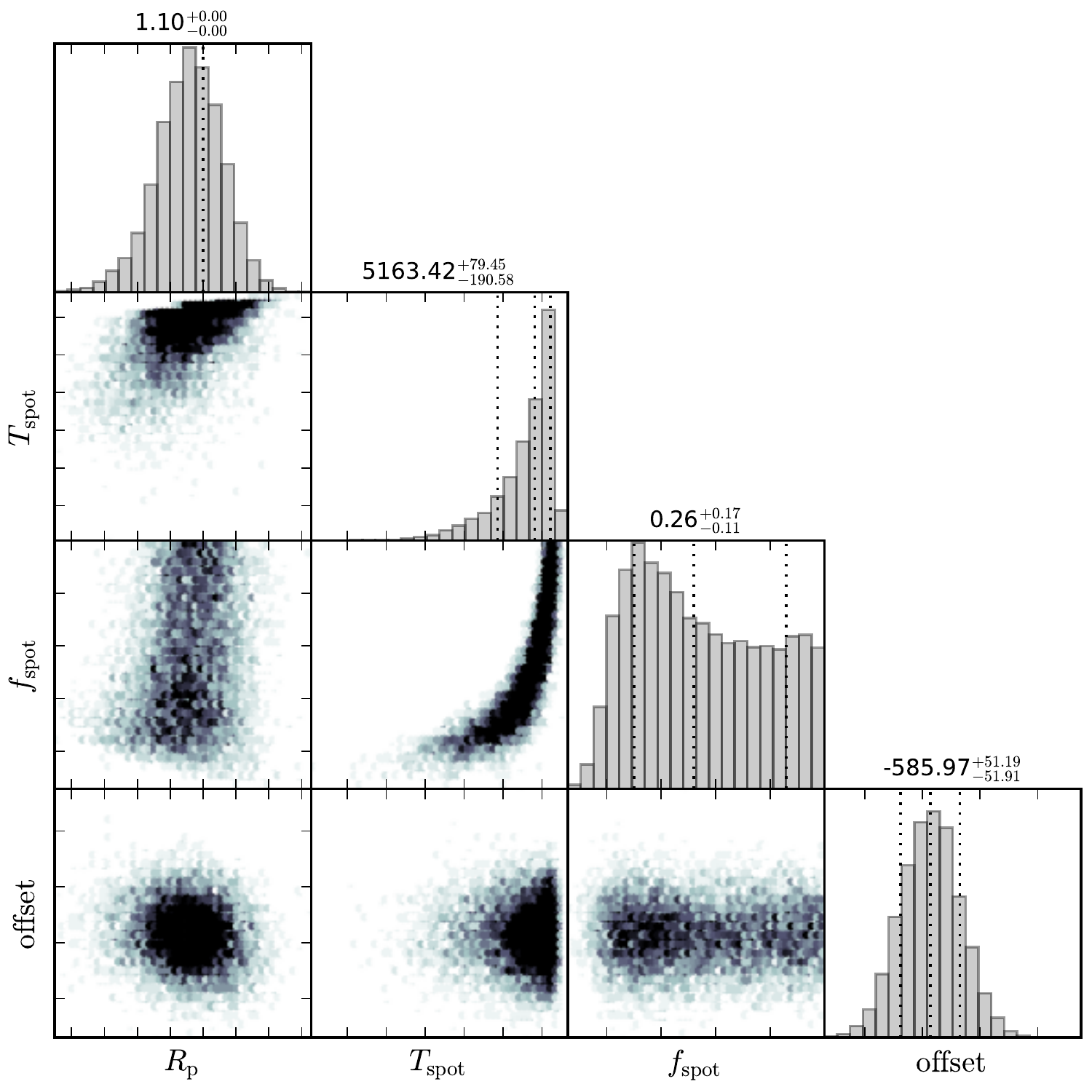}
\caption{Corner plot for the spectral retrieval analysis performed with \texttt{PLATON}, where spot contamination is the only slope origin.}
\label{fig:corner_platon_spot}
\end{figure}


\bsp	
\label{lastpage}
\end{document}